\documentclass[
aps,
 pra,
 amsmath,amssymb,
 reprint,%
longbibliography
]{revtex4-2}

\usepackage{graphicx}
\usepackage{dcolumn}
\usepackage{bm}

\usepackage{mathtools,slashed}
\usepackage{mathrsfs}

\usepackage[utf8]{inputenc}
\usepackage[T1]{fontenc}
\usepackage{etoolbox}
\usepackage{xr}
\usepackage{mathtools}

\graphicspath{
	{figures/}
}

\newcommand{\bk}[1][]{\mathbf{k}{#1}}
\newcommand{\br}[1][]{\mathbf{r}{#1}}

\newcommand{\ki}{\mathbf{k}_{\text{in}}}
\newcommand{\ks}{\mathbf{k}_{\text{s}}}

\newcommand{\sst}{_{\text{t}}}
\newcommand{\ssb}{_{\text{b}}}

\newcommand{\tsb}[1]{_{\text{#1}}}
\newcommand{\tsp}[1]{^{\text{#1}}}

\newcommand{\cop}[1][]{\hat{a}^{\dagger}_{\mathbf{k}{#1},\lambda{#1}}}
\newcommand{\aop}[1][]{\hat{a}_{\mathbf{k}{#1},\lambda{#1}}}
\newcommand{\be}[1][]{\bm{\varepsilon}_{\mathbf{k}_{#1},\lambda_{#1}}}

\newcommand{\ns}[1]{\{#1\}}

\begin{document}

\title{Unified approach to time-resolved x-ray and electron diffraction imaging}

\author{Mingrui Yuan}
\affiliation{Wyant College of Optical Sciences, University of Arizona, Tucson, AZ 85721}
\affiliation{Department of Physics, University of Arizona, Tucson, AZ 85721}

\author{Nikolay V. Golubev}
\email{ngolubev@arizona.edu}
\affiliation{Department of Physics, University of Arizona, Tucson, AZ 85721}

\date{\today}

\begin{abstract}
Time-resolved x-ray diffraction (TR-XRD) and ultrafast electron diffraction (TR-UED) are emerging tools for probing ultrafast quantum dynamics. From a theoretical perspective, they are commonly described within different frameworks and modeled using distinct approximations. Here, we present a unified quantum-field-based description of ultrafast diffraction imaging that permits consistent consideration of TR-XRD and TR-UED within a common theoretical formalism. Our approach elucidates the correspondence between TR-XRD and TR-UED and allows their similarities and differences to be systematically disentangled. The developed formalism is sufficiently general to consistently and straightforwardly incorporate additional physical effects of interest, such as relativistic charge-current and current-current couplings. We apply our approach to simulate diffraction measurements of laser-driven electron dynamics in graphene, demonstrating the unique capabilities of diffraction imaging to unravel intricate quantum processes in matter.
\end{abstract}

\pacs{Valid PACS appear here}
\maketitle

\section{Introduction}

The development of ultrafast laser technologies over the past several decades~\cite{krausz2009,krausz2016,hu2026} has fundamentally transformed our ability to probe and manipulate matter on its most elementary timescales. A broad range of experimental techniques, such as attosecond streaking~\cite{hentschel2001,mairesse2005,itatani2002}, reconstruction of attosecond beating by interference of two-photon transitions (RABBITT)~\cite{paul2001,mairesse2003}, high-harmonic spectroscopy~\cite{kraus2015}, time-resolved photoelectron~\cite{gruson2016} and photofragmentation~\cite{calegari2014,sansone2010} spectroscopies, attosecond transient absorption spectroscopy (ATAS)~\cite{goulielmakis2010,matselyukh2022,chakraborty2026}, to name just a few, has enabled the observation and control of ultrafast quantum dynamics in atoms~\cite{goulielmakis2010}, clusters~\cite{gong2022}, molecules~\cite{calegari2014,kraus2015,matselyukh2022}, liquids~\cite{jordan2020}, and solids~\cite{borrego-varillas2022,hui2022} with unprecedented resolution. Yet, while these methodologies excel at monitoring temporal evolution, they offer only limited insight into the spatial character of the underlying dynamics.

Real-space imaging of matter has traditionally been the domain of x-ray~\cite{cowley1995,schlichting2012} and electron~\cite{colliex2006,carter2016,hassan2017,hassan2018} diffraction, the foundational tools of atomic-scale investigations for more than a century. The ability to generate ultrashort x-ray pulses with modern free-electron lasers~\cite{emma2010,pellegrini2016,young2018} and ultrashort electron pulses via laser-triggered photoemission~\cite{fill2006,hommelhoff2006} and subsequent temporal compression~\cite{baum2007,hilbert2009,kealhofer2016,zheng2026} or gating~\cite{kozak2017,morimoto2018,morimoto2020,hui2024} has extended the diffraction imaging techniques deep into the ultrafast regime, enabling time-resolved diffraction measurements with femtosecond~\cite{yang2018a,wolf2019,liu2020a,yang2020,champenois2023,jiang2025} and sub-femtosecond~\cite{morimoto2018,hui2024} temporal resolution.

While the theoretical frameworks for describing x-ray and electron scattering from stationary targets are well established in the literature~\cite{coppens1992,coppens1997,cowley1993a,cowley1993b,santra2009}, their extension to time-resolved, non-equilibrium regimes remains an active area of research. Early attempts~\cite{cao1998a,authier2008,shao2010,baum2010,suominen2014,yakovlev2015,centurion2022} to directly extend time-independent scattering approaches to dynamical systems have been shown to be incomplete~\cite{santra2014} for describing diffraction in the ultrafast regime. It has become evident that a consistent description of time-resolved diffraction must explicitly account for both the coherent nature of the probe beam~\cite{robicheaux2000} and the interplay of elastic and inelastic scattering pathways it induces in the target~\cite{tanaka2001,bennett2014}.

In their pioneering work~\cite{dixit2012}, Dixit and coworkers developed fully quantum theory of the time-resolved x-ray diffraction (TR-XRD) that, together with efforts of other groups~\cite{bennett2014}, has laid the foundation for the modern theoretical description of ultrafast x-ray diffraction imaging~\cite{dixit2013a,dixit2013,biggs2014,dixit2014,popova-gorelova2015a,popova-gorelova2015b,rouxel2016,dixit2017,grosser2017,kowalewski2017,morenocarrascosa2017,bennett2018,cho2018,popova-gorelova2018a,simmermacher2019a,simmermacher2019,hermann2020,allum2021,giri2021,tremblay2021,giri2022,yong2022a,tremblay2023a,moghaddasifereidani2025,radionov2025,moghaddasifereidani2026}.  In parallel, Shao and Starace developed a theory of time-resolved ultrafast electron diffraction (TR-UED)~\cite{shao2013,shao2013a,shao2014,shao2016,shao2017}. In contrast to the TR-XRD framework of Dixit and co-workers, which is based on time-dependent perturbation theory, Shao and Starace employed the Lippmann--Schwinger equation to derive the corresponding expression for the scattering cross section in TR-UED. Although the resulting expressions in TR-XRD and TR-UED share certain similarities, a direct comparison between them, to the best of our knowledge, has not yet been carried out, and it therefore remains unclear how closely the TR-XRD and TR-UED approaches are related to each other.

In recent years, a number of studies~\cite{rouxel2021,yong2022b,yong2022,yong2023,sun2024,wu2025,yuan2025} have reported the formal use of TR-XRD expressions in the context of electron diffraction. While there are many analogies between x-ray and electron scattering, and the corresponding time-independent cross sections become identical up to a prefactor within certain approximations, an \textit{ad hoc} replacement of equations between TR-XRD and TR-UED is not generally justified, as it may overlook important differences in their underlying dynamical formulations and the assumptions entering the respective derivations.

In this study, we report a unified quantum-field-based theory of ultrafast diffraction imaging that permits consistent consideration of TR-XRD and TR-UED within a common theoretical framework. Starting from the time-dependent formulation of x-ray diffraction imaging reported by Dixit and co-workers~\cite{dixit2012}, we extend the framework to the scattering of arbitrary particles, thereby providing a unified description that naturally connects with the formalism developed by Shao and Starace for electron diffraction~\cite{shao2013,shao2014}. Within the developed theory, both probe types are naturally encompassed within a single consistent methodology, enabling a transparent identification of the assumptions underlying each description and allowing the similarities and differences between TR-XRD and TR-UED to be made explicit. The resulting formulation provides a flexible starting point for systematically incorporating additional physical effects, including various relativistic corrections, thereby broadening the scope of ultrafast diffraction theory. While these effects are generally of lower importance in the physics of the time-resolved diffraction measurements, we show that under certain conditions they can play a leading role in the formation of the corresponding signal. We illustrate the applicability of our approach by simulating diffraction measurements of laser-driven electron dynamics in graphene, highlighting how diffraction imaging can provide detailed insight into complex quantum processes in matter and access their underlying physical mechanisms.

The remainder of this paper is organized as follows. In Sec.~\ref{sec:scattering}, we briefly review the general scattering formalism and the associated observables, followed in Sec.~\ref{sec:system} by a fully quantum-mechanical description of both the target and the probe beam. In Secs.~\ref{sec:stationary_formalism} and~\ref{sec:td_formalism}, we present expressions for the time-independent and time-dependent, respectively, scattering probabilities. We demonstrate in Sec.~\ref{sec:LippSchwin} the connection between the time-dependent formulation of scattering theory and the corresponding energy-domain Lippmann--Schwinger formalism. Sections~\ref{sec:Xray_st} and~\ref{sec:Xray_td} present explicit derivations of x-ray scattering probabilities for the stationary and time-dependent cases, respectively, while Secs.~\ref{sec:el_st} and~\ref{sec:td_el_nonrel} introduce the corresponding stationary and time-dependent scattering probabilities for electrons. In Sec.~\ref{sec:relel}, we present the general expression for the electron scattering probability that takes into account relativistic charge-current and current-current couplings. We discuss the resulting expressions, as well as their similarities and differences, in Sec.~\ref{sec:discuss}. In Sec.~\ref{sec:applications}, we present applications of the developed techniques to the case of laser-driven electron dynamics in monolayer graphene sample. In Sec.~\ref{sec:concl}, we summarize the results and conclude this paper. In addition, Appendices~\ref{app:xray_beam} and~\ref{app:el_beam} provide a detailed discussion of the fully quantum-mechanical description of x-ray and electron beams, respectively, and their interactions with a target. Appendix~\ref{app:rel_beam} presents the corresponding formulation of relativistic electron-electron interactions, while some of the expressions employed therein are derived in Appendix~\ref{app:contraction}.

\section{Theoretical framework}

\subsection{Scattering probability and cross-section}
\label{sec:scattering}

Let us consider a general case of a beam interacting with a target. We denote by $P(\ks)$ the probability of observing a scattered beam with momentum $\ks$, which is different from the momentum of the incoming beam. The total scattering probability $P$ is then obtained by summing over all possible values of the scattered momentum $\ks$:
\begin{equation}
\label{eq:P_sum}
    P = \sum_{\ks} P(\ks).
\end{equation}
This probability characterizes the ratio between the number $N_{\text{s}}$ of the scattered particles and the total number $N_{\text{in}}$ of incident beam particles that strike the target during the course of the experiment:
\begin{equation}
\label{eq:P_particles}
	P=N_{\text{s}}/N_{\text{in}}.
\end{equation}

In the continuum limit, the sum in Eq.~(\ref{eq:P_sum}) can be replaced by an integral according to $\sum_{\ks} \to V/(2\pi)^3 \int d^3 \ks$ relation, where $V$ is the quantization volume. Expressing the volume element in spherical coordinates as $d^3\ks=|\ks|^2 d|\ks|d\Omega$, where $d\Omega$ is a small solid angle in the direction ($\theta$, $\phi$), the total scattering probability becomes
\begin{equation}
\label{eq:total_P}
    P = \frac{V}{(2\pi)^3} \int P(\ks) |\ks|^2 d|\ks|d\Omega.
\end{equation}

Since the scattering probability depends on experimental conditions like target density, thickness, and beam intensity, it is, in some cases, more convenient to utilize the cross section, which provides a universal and geometry-independent measure of the intrinsic interaction between the beam and the target. The scattering cross section $\sigma$ can be defined as an effective area that relates the number of scattered particles to the time-dependent beam flux $F(t)$ integrated over the full duration of the measurement:
\begin{equation}
\label{eq:sigma_general}
	\sigma = \frac{N_{\text{s}}}{\int_{t_i}^{t_f} F(t) dt},
\end{equation}
where $t_i$ and $t_f$ denote the initial and final times, respectively. Eq.~(\ref{eq:sigma_general}) extends the standard definition of the scattering cross section (see, e.g., Eq.~(5.2) in Ref.~\cite{schwartz2014}) to situations involving a time-dependent beam. Comparing Eqs.~(\ref{eq:P_particles}) and~(\ref{eq:sigma_general}), one can see that the scattering probability is related to the scattering cross section as
\begin{equation}
\label{eq:sigma_P}
    \sigma = P \frac{N_{\text{in}}}{\int_{t_i}^{t_f} F(t) dt}.
\end{equation}

To characterize how the scattering probability and cross section are distributed with respect to the direction of the scattered momentum, it is convenient to introduce the differential scattering probability and the associated differential scattering cross section. The former can be obtained from Eq.~(\ref{eq:total_P}) by differentiating the total probability with respect to the solid angle $\Omega$:
\begin{equation}
\label{eq:dP}
    \frac{dP}{d\Omega} = \frac{V}{(2\pi)^3} \int_0^{+\infty} P(\ks) |\ks|^2 d|\ks|,
\end{equation}
and the corresponding differential scattering cross section can then be obtained using the relation in Eq.~(\ref{eq:sigma_P}). Furthermore, it can be convenient to replace the integration over the momentum magnitude in Eq.~(\ref{eq:dP}) with an integration over energy. Making the substitution $d|\ks|=\frac{d|\ks|}{dE\tsb{s}} dE\tsb{s}$, we obtain
\begin{equation}
    \frac{dP}{d\Omega} = \frac{V}{(2\pi)^3} \int_0^{+\infty} P(\ks) |\ks|^2 \frac{d|\ks|}{dE\tsb{s}} dE\tsb{s},
\end{equation}
which, together with Eq.~(\ref{eq:dP}) and the corresponding expressions for the differential scattering cross sections, constitutes the basis for the subsequent analysis of both x-ray and electron scattering.

\subsection{Description of the system}
\label{sec:system}

The total Hamiltonian of the system can be written as
\begin{equation}
\label{eq:H_total}
    \hat{H} = \hat{H}_0 + \hat{H}_{\text{int}},
\end{equation}
where
\begin{equation} 
    \hat{H}_0 = \hat{H}_{\text{t}} + \hat{H}_{\text{b}},
\end{equation}
describes the unperturbed system composed of the respective Hamiltonians of the target $\hat{H}_{\text{t}}$ and the probe beam $\hat{H}_{\text{b}}$, and $\hat{H}_{\text{int}}$ is the interaction Hamiltonian between them. These operators can, in principle, be time dependent, allowing one to describe scattering from a target explicitly driven by interactions with external fields, or to address more general situations such as the dynamical modification of the beam during its propagation.

The dynamics of the system described by the Hamiltonian Eq.~(\ref{eq:H_total}) in time is governed by the evolution operator $\hat{U}(t_f,t_i)$. Applying the time-dependent perturbation theory with respect to the interaction Hamiltonian $\hat{H}_{\text{int}}$, one can write:
%
\begin{equation}
\label{eq:U}
    \hat{U}(t_f,t_i) = \hat{U}_0(t_f,t_i)
	- i \int_{t_i}^{t_f} dt\ \hat{U}_0(t_f,t) \hat{H}_{\text{int}} \hat{U}_0(t,t_i)
        + ...,
\end{equation}
%
where $\hat{U}_0(t_f,t_i)$ is the evolution operator associated with Hamiltonian $\hat{H}_0$. Since the unperturbed Hamiltonian $\hat{H}_0$ is composed of contributions from distinct subsystems, it is natural to represent the Hilbert space of the full system as a tensor product $\mathcal{H}=\mathcal{H}_{\text{t}} \otimes \mathcal{H}_{\text{b}}$. Accordingly, the unperturbed evolution operator can be factorized as
\begin{equation}
    \hat{U}_0(t_f,t_i) = \hat{U}\tsb{t}(t_f,t_i) \hat{U}\tsb{b}(t_f,t_i),
\end{equation}
where $\hat{U}\tsb{t}(t_f,t_i)$ and $\hat{U}\tsb{b}(t_f,t_i)$ are the evolution operators for the target and the beam, respectively. Therefore, in the absence of interactions, the target and the probe beam evolve independently. The interaction Hamiltonian $\hat{H}_{\text{int}}$ couples the unperturbed states of the target and the beam and drives transitions between them, thereby governing the time evolution of the total system during the scattering process.

In the absence of external fields, an eigenstate $|\Psi_I\rangle$ of the unperturbed system can be expressed in the tensor-product form as
\begin{equation}
	|\Psi_I\rangle = |\Phi_i \rangle \otimes |\psi_j \rangle 
		\equiv |\Phi_i \rangle |\psi_j \rangle,
\end{equation}
where $|\Phi_i \rangle$ and $|\psi_j \rangle$ are eigenstates of $\hat{H}\sst$ and $\hat{H}\ssb$, respectively:
\begin{eqnarray}
	\hat{H}\tsb{t} & |\Phi_i\rangle  = E_i\tsp{t} & |\Phi_i\rangle, \\
	\hat{H}\tsb{b} & |\psi_j\rangle  = E_j\tsp{b} & |\psi_j\rangle,
\end{eqnarray}
with $E_i\tsp{t}$ and $E_j\tsp{b}$ denoting the corresponding eigenenergies. The eigenenergy $E_I$ of the total system is
\begin{equation}
	E_I = E_i\tsp{t} + E_j\tsp{b},
\end{equation}
which arises due to a separable structure of $\hat{H}_0$ Hamiltonian. Under these assumptions, the field-free evolution operators of the target and the beam can be written as
\begin{eqnarray}
    \hat{U}\tsb{t}(t_f,t_i) = \sum_i e^{-i E_i\tsp{t} (t_f - t_i)} |\Phi_i\rangle \langle \Phi_i|, \\
    \hat{U}\tsb{b}(t_f,t_i) = \sum_j e^{-i E_j\tsp{b} (t_f - t_i)} |\psi_j\rangle \langle \psi_j|.
\end{eqnarray}

While the eigenstates $|\psi_j\rangle$ provide a natural description of the time evolution of the beam, they are not necessarily the most convenient choice for representing the beam in scattering problems. Indeed, in a typical scattering experiment one is primarily interested to measure the change in the number of particles occupying certain momentum modes characterized by the corresponding wave vectors $\bk_i$. In this context, the Fock number state basis built from single-particle momentum eigenstates offers a more suitable representation.

For a given mode $i$ with wave vector $\bk_i$, the single-mode number state $|n_i\rangle$ is defined as the eigenstate of the number operator $\hat{N}_i= \hat{a}_i^\dagger \hat{a}_i$, such that
\begin{equation}
    \hat{N}_i |n_i\rangle = \hat{a}_i^\dagger \hat{a}_i |n_i\rangle = n_i |n_i\rangle, 
\end{equation}
where $n_i$ is the number of particles present in the corresponding mode, and $\hat{a}_i^\dagger$ and $\hat{a}_i$ are the creation and annihilation operators, respectively. 

The many-body number states are given by
\begin{equation}
    |\{n\}\rangle = \bigotimes_{j} |n_j\rangle=
    |n_1, n_2, \dots,n_N \rangle,
\end{equation}
where $n_j$ denotes the occupation number of the $j$-th mode. These states form a complete orthonormal basis of the many-body Hilbert space, satisfying
\begin{equation}
    \langle \{\bar{n}\} | \{n\} \rangle = \prod_{j=1}^{N} \delta_{\bar{n}_j, n_j},
\end{equation}
and the completeness relation is
\begin{equation}
\sum_{\{n\}} |\{n\}\rangle \langle \{n\}| = \mathbb{I}.
\end{equation}

In the basis of many-body number states, the creation $\hat{a}_i^{\dagger}$ and annihilation $\hat{a}_j$ operators obey either commutation or anticommutation relations, depending on whether the particles are photons or electrons, corresponding to bosonic or fermionic statistics, respectively. The explicit forms of these operators for these two cases are given in Appendices~\ref{app:xray_beam} and~\ref{app:el_beam}. In general, the representation of the beam in terms of many-body number states, together with the associated field operators, provides a convenient and complete framework for describing systems with variable particle number~\cite{mandel2008,fetter2012}, and will be employed throughout this work to represent the beam and evaluate observables.

\subsection{Elastic scattering from a stationary target}
\label{sec:stationary_formalism}

The scattering theory of electrons and x-rays from a stationary target is well developed in the literature (see, e.g., Refs.~\cite{newton1982,goldberger2004,sakurai2020}). Here, we briefly outline the main concepts of time-independent scattering in order to establish the notation and provide a basis for comparison with the time-dependent formulations discussed later.

In mathematical terms, elastic scattering from a stationary target corresponds to the target being present in an energy eigenstate $|\Phi_j \rangle$, most commonly the ground state, of the Hamiltonian $\hat{H}\tsb{t}$ both before and after the scattering event. We assume a continuous incoming beam with well-defined momentum $\ki$, described by the single-mode number state $|n_{\ki}\rangle$. The scattered beam, in turn, is described by a general number state $|\{m\}\rangle$ which contains particles in the momentum mode $\ks$. For the sake of simplicity, we assume that both the initial and the scattered states of the beam are the eigenstates of the beam Hamiltonian, such that $|n_{\ki}\rangle=|\psi^{\ki}_i \rangle$ and $|\ns{m}\rangle=|\psi^{\ks}_f \rangle$.

The scattering probability $P(\ks)$ to find the beam in the state $|\psi^{\ks}_f \rangle$ can be obtained as
\begin{equation}
	P(\ks) = \left| S_{FI} \right|^2,
\end{equation}
where $S_{FI}$ is the amplitude of transition from the initial state $|\Psi_I\rangle=|\Phi_j \rangle |\psi^{\ki}_i \rangle$ of the system long before the collision to the state $|\Psi_F\rangle=|\Phi_j \rangle |\psi^{\ks}_f \rangle$ long after the collision. 

To evaluate the transition amplitude $S_{FI}$, the initial state $|\Psi_I\rangle$ is propagated under the action of the full Hamiltonian $\hat{H}$ and projected onto the desired final state $|\Psi_F\rangle$:
\begin{equation}
	S_{FI} = \lim_{\substack{t_i \to -\infty \\ t_f\to+\infty}} 
		\langle \Psi_F | \hat{U}(t_f,t_i) | \Psi_I \rangle.
\end{equation}
Using the perturbative expansion of $\hat{U}(t_f,t_i)$, Eq.~(\ref{eq:U}), up to the first order, one can express the transition amplitude as
\begin{equation}
\label{eq:SFI_st}
	S_{FI} = -2\pi i \delta(E\tsp{b}_f-E\tsp{b}_i) \langle \Phi_j | \langle \psi^{\ks}_f | \hat{H}_{\text{int}} |\psi^{\ki}_i \rangle | \Phi_j \rangle,
\end{equation}
where the zeroth order contribution vanished completely due to the orthogonality of $|\psi^{\ki}_i \rangle$ and $|\psi^{\ks}_f \rangle$ states and we utilized the fact that $E_F - E_I = E\tsp{b}_f-E\tsp{b}_i$ since the target remains in the same eigenstate.

Using the standard replacement for the squared delta function in a finite interaction time $T$, $\delta^2(E)=\frac{T}{2\pi}\,\delta(E)$~(see, e.g., Ref.~\cite{santra2009}), the scattering probability can be written as
\begin{equation}
\label{eq:Pks_stationary}
	P(\ks) = 2\pi T \delta(E\tsp{b}_f-E\tsp{b}_i) 
		\left |
			\langle \Phi_j | \langle \psi^{\ks}_f | \hat{H}_{\text{int}} |\psi^{\ki}_i \rangle | \Phi_j \rangle
		\right |^2.
\end{equation}
This expression shows that the elastic scattering corresponds to a transition between beam eigenstates induced by the interaction with the target that remains in the same internal state, with conservation of beam energy enforced by the Dirac delta function. The matrix element $\langle \Phi_j | \langle \psi^{\ks}_f | \hat{H}_{\text{int}} |\psi^{\ki}_i \rangle | \Phi_j \rangle$ therefore fully determines the probability of scattering to the direction defined by the momentum $\ks$.

\subsection{Scattering from a time-dependent target}
\label{sec:td_formalism}

The quantum theory of scattering from a time-dependent target was explicitly formulated in Ref.~\cite{dixit2012}. In that study and subsequent works~\cite{dixit2013a,dixit2013,dixit2014,dixit2017,hermann2020,giri2021,tremblay2021,giri2022,tremblay2023a}, the authors focused exclusively on x-ray scattering. In this section, we present a generalized version of their approach that is independent of the nature of the probe beam.

In the time-dependent case, it is convenient to employ the density-matrix formalism, which treats the target and the beam on an equal footing and naturally allows for mixed states, thereby enabling a more flexible description of the beam. The total system is described by the density matrix $\hat{\rho}(t)$, which time evolution can be written as
\begin{equation}
    \hat{\rho}(t) =
        \hat{U}(t,t_i) \hat{\rho}(t_i) \hat{U}^{\dagger}(t,t_i).
\end{equation}
At the initial moment of time $t_i$, the total density matrix can be constructed from the density matrices of the target and the beam as
\begin{equation}
    \hat{\rho}(t_i)=\hat{\rho}\tsb{t}(t_i) \otimes \hat{\rho}\tsb{b}(t_i).
\end{equation}
Although the initial density matrix factorizes into independent beam and target contributions, their interaction during the time evolution generally leads to the formation of correlations between the two subsystems. Consequently, at later times the total density matrix cannot, in general, be expressed as a simple product of the corresponding subsystem density matrices. 

We assume the target is present in a pure state which at the initial moment of time is given by the following density matrix
\begin{equation}
\label{eq:rho_target}
    \hat{\rho}\tsb{t}(t_i) = |\Phi(t_i)\rangle \langle \Phi(t_i)|,
\end{equation}
where the corresponding initial wavefunction is given by
\begin{equation}
    |\Phi(t_i)\rangle = \sum_{i} C_i|\Phi_i\rangle,
\end{equation}
and $C_i$ are, in general complex, expansion coefficients.

The initial state of the beam is described by a general density matrix
\begin{equation}
\label{eq:beam_density}
    \hat{\rho}\tsb{b}(t_i) = \sum_{\{n\},\{\bar{n}\}} \rho\tsp{b}_{\{n\},\{\bar{n}\}}|\{n\}\rangle \langle \{\bar{n}\}|,
\end{equation}
which, depending on the choice of matrix elements $\rho\tsp{b}_{\{n\},\{\bar{n}\}}$, may represent either pure or mixed state of the beam. Importantly, the number state basis used in the representation of the density matrix in Eq.~(\ref{eq:beam_density}) is complete. Physically relevant beam configurations, for example those corresponding to an initial beam propagating along the $\ki$ direction, are obtained through an appropriate choice of the matrix elements $\rho\tsp{b}_{\{n\},\{\bar{n}\}}$, which encode the occupation and coherence properties of the corresponding momentum modes.

Combining the expressions for the target and beam density matrices, the initial density matrix of the entire system can be written as
\begin{equation}
    \hat{\rho}(t_i)= \sum_{\{n\},\{\bar{n}\}} \rho\tsp{b}_{\{n\},\{\bar{n}\}} |\{n\} \rangle \langle \{\bar{n}\}| \otimes |\Phi(t_i)\rangle \langle \Phi(t_i)|.
\end{equation}
In principle, a fully general density matrix of the system, constructed from tensor products $|\ns{n}\rangle \otimes |\Phi_i\rangle$, can also be prepared and used in the subsequent analysis, thereby allowing for the treatment of mixed states of the target. However, for clarity and without loss of generality for the present discussion, we restrict ourselves to a description of the target in the pure state, noting that the extension to mixed states follows straightforwardly within the same formalism.

The probability of detecting a scattered beam with momentum $\ks$ can be defined as
\begin{equation}
    P(\ks)=\text{Tr}\left[ \hat{O}_{\ks} \hat{\rho}(t_f)\right],
\end{equation}
where the operator $\hat{O}_{\ks}$ is given by
\begin{equation}
    \hat{O}_{\ks}=\sum_l \sum_{\{m\}} 
        |\Phi_l\rangle \langle \Phi_l| \otimes
        |\{m\}\rangle \langle\{m\}|.
\end{equation}
This operator projects the density matrix of the system at the final moment of time, $\hat{\rho}(t_f) \equiv \hat{\rho}(t \to +\infty)$, onto the states $|\{m\}\rangle$ in which the beam contains particles with the momentum $\ks$. The target, in turn, is allowed to occupy any of its possible states $|\Phi_l\rangle$, so that the trace effectively sums over all final configurations of the target.

Combining the above expressions together, the scattering probability can be written as
\begin{widetext}
\begin{equation}
    P(\ks)=\lim_{\substack{t_i \to -\infty \\ t_f\to+\infty}} \sum_{l}\sum_{\{m\}} \sum_{\{n\},\{\bar{n}\}} \rho\tsp{b}_{\{n\},\{\bar{n}\}}
        \langle \Phi_l| \langle \ns{m}|  \hat{U}(t_f,t_i)  |\ns{n}\rangle |\Phi(t_i)\rangle
        \langle \Phi(t_i)| \langle \ns{\bar{n}} | \hat{U}^{\dagger}(t_f,t_i) |\ns{m} \rangle |\Phi_l \rangle.
\end{equation}
%
Exchanging the order of terms in this expression and utilizing the resolution of identity $\sum_{l} |\Phi_l \rangle\langle \Phi_l| \equiv \mathbb{I}$ for the target states, we obtain
%
\begin{equation}
\label{eq:Pks_general}
    P(\ks)=\lim_{\substack{t_i \to -\infty \\ t_f\to+\infty}} \sum_{\ns{m}} \sum_{\ns{n},\ns{\bar{n}}} \rho\tsp{b}_{\ns{n},\ns{\bar{n}}}
        \langle \Phi(t_i)| \langle \ns{\bar{n}} | \hat{U}^{\dagger}(t_f,t_i) |\ns{m} \rangle
        \langle \ns{m}|  \hat{U}(t_f,t_i) |\ns{n}\rangle |\Phi(t_i)\rangle.
\end{equation}
%
To proceed forward, we use the perturbative expansion of $\hat{U}(t_f,t_i)$ operator, Eq.~(\ref{eq:U}), retaining terms up to first order in the interaction Hamiltonian, to express the corresponding matrix elements:
\begin{equation}
\label{eq:beam_Udag}
    \langle \ns{\bar{n}}|  \hat{U}^{\dagger}(t_f,t_i) |\ns{m}\rangle =
    i  \int_{t_i}^{t_f} dt_2\
    \hat{U}^{\dagger}\tsb{t}(t_2,t_i)
        \langle \ns{\bar{n}}| \hat{U}^{\dagger}\tsb{b}(t_2,t_i)
        \hat{H}^{\dagger}_{\text{int}} 
        \hat{U}^{\dagger}\tsb{b}(t_f,t_2) |\ns{m} \rangle
    \hat{U}^{\dagger}\tsb{t}(t_f,t_2)
\end{equation}
%
%
%
and
\begin{equation}
\label{eq:beam_U}
    \langle \ns{m}|  \hat{U}(t_f,t_i) |\ns{n}\rangle = 
        -i  \int_{t_i}^{t_f} dt_1\
            \hat{U}\tsb{t}(t_f,t_1)
                 \langle \ns{m}| \hat{U}\tsb{b}(t_f,t_1)
                    \hat{H}_{\text{int}} 
                 \hat{U}\tsb{b}(t_1,t_i) |\ns{n} \rangle
            \hat{U}\tsb{t}(t_1,t_i).
\end{equation}
In principle, since $|\ns{n}\rangle$ and $|\ns{\bar{n}}\rangle$ form complete sets that include the states $|\ns{m}\rangle$, the corresponding matrix elements Eqs.~(\ref{eq:beam_Udag}) and~(\ref{eq:beam_U}) should also contain zeroth-order contributions which, in contrast to Eq.~(\ref{eq:SFI_st}), do not vanish when $|\ns{n}\rangle=|\ns{m}\rangle$ or $|\ns{\bar{n}}\rangle=|\ns{m}\rangle$. However, when focusing on the scattering problems, such that $\ks \neq \ki$, these contributions are excluded from consideration by restricting the density matrix, i.e., setting $\rho\tsp{b}_{\{n\},\{\bar{n}\}}=0$ for those $|\ns{n}\rangle$ and $|\ns{\bar{n}}\rangle$ that coincide with $|\ns{m}\rangle$.

%

Substituting Eqs.~(\ref{eq:beam_Udag}) and~(\ref{eq:beam_U}) to Eq.~(\ref{eq:Pks_general}) and using the relation $\hat{U}^{\dagger}\tsb{t}(t_f,t_2)\hat{U}\tsb{t}(t_f,t_1)=\hat{U}\tsb{t}(t_2,t_1)$, the scattering probability can be expressed as
\begin{equation}
\begin{split}
\label{eq:General_Scattering_Prob}
    P(\ks)=\lim_{\substack{t_i \to -\infty \\ t_f\to+\infty}}  
        & \sum_{\ns{m}} \sum_{\ns{n},\ns{\bar{n}}}   
        \rho\tsp{b}_{\ns{n},\ns{\bar{n}}}  
        \int_{t_i}^{t_f} dt_1 \int_{t_i}^{t_f} dt_2  \\
        \times & 
        \langle \Phi(t_2)|
            \langle \ns{\bar{n}}| 
                \hat{U}^{\dagger}\tsb{b}(t_2,t_i) \hat{H}^{\dagger}_{\text{int}} \hat{U}^{\dagger}\tsb{b}(t_f,t_2) 
            |\ns{m} \rangle
        \hat{U}\tsb{t}(t_2,t_1) 
            \langle \ns{m}| 
                \hat{U}\tsb{b}(t_f,t_1) \hat{H}_{\text{int}} \hat{U}\tsb{b}(t_1,t_i)  
            |\ns{n} \rangle
        |\Phi(t_1)\rangle.
\end{split}
\end{equation}
\end{widetext}
It is seen that the scattering probability reflects the evolution of the target in time through the interference of transitions from initial beam states $|\ns{n} \rangle$ and $|\ns{\bar{n}} \rangle$ into common final states $|\ns{m} \rangle$. The resulting expression depends on the matrix elements $\langle \ns{m}| \hat{U}\tsb{b}(t_f,t_1) \hat{H}_{\text{int}} \hat{U}\tsb{b}(t_1,t_i) |\ns{n} \rangle$ and $\langle \ns{\bar{n}}| \hat{U}^{\dagger}\tsb{b}(t_2,t_i) \hat{H}^{\dagger}_{\text{int}} \hat{U}^{\dagger}\tsb{b}(t_f,t_2) |\ns{m} \rangle$, which are fully determined by the form of the evolution operator $\hat{U}\tsb{b}$ and the specific nature of the interaction Hamiltonian $\hat{H}_{\text{int}}$.

\subsection{Lippmann--Schwinger formalism for description of time-dependent scattering}
\label{sec:LippSchwin}

In parallel with the development of time-dependent formalisms for describing scattering from an evolving target, an alternative approach based on the Lippmann--Schwinger equation and employing Born series has been proposed~\cite{shao2013,shao2014}. In a difference to Ref.~\cite{dixit2012}, where the scattering of x-rays has been considered, the authors of Refs.~\cite{shao2013,shao2014} constructed their framework in application to the electron scattering. While the obtained equations for x-ray and electron scattering share certain similarities, it is not immediately clear how closely the two approaches are related or whether time-dependent scattering of x-rays and electrons probe different aspects of the target dynamics.

In this section, we elucidate the connection between the approaches presented in Refs.~\cite{dixit2012} and~\cite{shao2013,shao2014} and demonstrate that they yield mutually consistent results. Indeed, the time-dependent perturbation theory used in the previous section to derive the scattering probability is expected to reproduce the Born approximation, reflecting the close correspondence between the Dyson series in the time domain and the Born series in the energy domain. To make the comparison more transparent, we adopt in this section, where appropriate, the notation used in Ref.~\cite{shao2014}.

Starting from the density matrix description of the beam, Eq.~(\ref{eq:beam_density}), we restrict our analysis to a beam of non-interacting particles prepared in a pure state expressed in terms of single-mode states $|\bk_0\rangle = |n_{\bk_0}\rangle$ such that
\begin{equation}
    \hat{\rho}\tsb{b}(t_i) = 
    \int d \mathbf{k}_{0}^\prime
    \int d \mathbf{k}_{0} ~
    a^*_0(\mathbf{k}_{0}^\prime)
    a_0(\mathbf{k}_{0})
    |\mathbf{k}_{0}\rangle 
    \langle \mathbf{k}_{0}^\prime |,
\end{equation}
where $a_0(\mathbf{k}_{0})$ are the momentum amplitudes specifying a continuous momentum distribution of the beam.

The target is assumed to be in a pure state, moving in space, and prepared in a superposition of internal states, such that the initial density matrix of the target, Eq.~(\ref{eq:rho_target}), can be written as
\begin{widetext}
\begin{equation}
    \hat{\rho}\tsb{t}(t_i) = 
    \sum_{n^\prime} \sum_{n}
    C_{n^\prime}^* C_n
    \int d \mathbf{k}_{1}^\prime 
    \int d \mathbf{k}_{1} ~
    a^*_1(\mathbf{k}_{1}^\prime)
    a_1(\mathbf{k}_{1})
    |n,\mathbf{k}_{1}\rangle 
    \langle n^\prime, \mathbf{k}_{1}^\prime |,
\end{equation}
where $|n,\mathbf{k}_{1}\rangle$ labels the combined states of the target, with $n$ denoting the internal eigenstate $|\Phi_n\rangle$ and $\mathbf{k}_{1}$ the momentum, and $C_n$ and $a_1(\mathbf{k}_{1})$ are the corresponding expansion coefficients and momentum amplitudes, respectively.

Combining the expressions for the target and beam density matrices, the initial density matrix of the entire system will be
\begin{equation}
    \hat{\rho}(t_i)=     
    \sum_{n^\prime}\sum_{n}
    C_{n^\prime}^* C_n
    \int d \mathbf{k}'_{0} 
    \int d \mathbf{k}_{0} ~
    \int d \mathbf{k}'_{1} 
    \int d \mathbf{k}_{1} ~
    a^*_0(\mathbf{k}_{0}^\prime)
    a_0(\mathbf{k}_{0}) 
    a^*_1(\mathbf{k}_{1}^\prime) 
    a_1(\mathbf{k}_{1}) ~
    |n,\mathbf{k}_{1}\rangle 
    \langle n^\prime, \mathbf{k}_{1}^\prime |
    \otimes |\mathbf{k}_{0}\rangle 
    \langle \mathbf{k}_{0}^\prime |.
\end{equation}
The operator describing the probability of observing the beam in $|\ks\rangle$ state while the target is present in an arbitrary internal and momentum state $|l, \mathbf{k}_{b} \rangle$ can be written as
\begin{equation}
    \hat{O}_{\ks}
    =
    \sum_l
    \int d\mathbf{k}_b ~
    |\ks\rangle \langle\ks|
    \otimes
    |l, \mathbf{k}_{b} \rangle 
    \langle l, \mathbf{k}_{b}|.
\end{equation}
Therefore, the corresponding scattering probability can be expressed as
\begin{equation}
\label{eq:P_continuus}
\begin{split}
    P(\ks)=
    \sum_l P_l(\ks) =
    \lim_{\substack{t_i \to -\infty \\ t_f\to+\infty}}
    &
    \sum_l
    \int d\mathbf{k}_b
    \sum_{n'}\sum_{n}
    C_{n'}^* C_{n}
    \int d \mathbf{k}'_{0}
    \int d \mathbf{k}_{0} ~
    \int d \mathbf{k}'_{1} 
    \int d \mathbf{k}_{1} ~
    a^*_0(\mathbf{k}_{0}^\prime)
    a_0(\mathbf{k}_{0}) 
    a^*_1(\mathbf{k}_{1}^\prime) 
    a_1(\mathbf{k}_{1})
    \\
    \times & 
    \Big[
        \langle l, \mathbf{k}_{b}|
        \otimes 
        \langle\ks| 
    \Big]
    \hat{U}(t_f,t_i)
    \Big[
        |\mathbf{k}_{0}\rangle 
        \otimes 
        |n,\mathbf{k}_{1}\rangle 
    \Big]
    \Big[
        \langle n^\prime, \mathbf{k}_{1}^\prime |
        \otimes 
        \langle \mathbf{k}_{0}^\prime |
    \Big]
    \hat{U}^{\dagger}(t_f,t_i)
    \Big[
        |\ks\rangle
        \otimes
        |l, \mathbf{k}_{b} \rangle 
    \Big],
\end{split}
\end{equation}
where $P_l(\ks)$ denotes the partial transition probability to find the beam in $|\ks\rangle$ state while the target is present in the internal state $l$.

The transition matrix elements present in Eq.~(\ref{eq:P_continuus}) can be recognized as the elements of the scattering $S$-matrix such that
\begin{equation}
    \lim_{\substack{t_i \to -\infty \\ t_f\to+\infty}}
    \Big[
        \langle l, \mathbf{k}_{b}|
        \otimes 
        \langle\ks| 
    \Big]
    \hat{U}(t_f,t_i)
    \Big[
        |\mathbf{k}_{0}\rangle 
        \otimes 
        |n,\mathbf{k}_{1}\rangle 
    \Big]
     = 
     \langle f | \hat{S} |i \rangle = S_{fi},
\end{equation}
and 
\begin{equation}
    \lim_{\substack{t_i \to -\infty \\ t_f\to+\infty}}
    \Big[
        \langle n^\prime, \mathbf{k}_{1}^\prime |
        \otimes 
        \langle \mathbf{k}_{0}^\prime |
    \Big]
    \hat{U}^{\dagger}(t_f,t_i)
    \Big[
        |\ks\rangle
        \otimes
        |l, \mathbf{k}_{b} \rangle
    \Big]
     = 
     \langle i' | \hat{S}^{\dagger} |f \rangle = S^{*}_{fi'},
\end{equation}
where we introduced the compact notation $|i\rangle \equiv |\mathbf{k}_{0}\rangle \otimes |n,\mathbf{k}_{1}\rangle$, $|i' \rangle \equiv |\mathbf{k}'_{0}\rangle \otimes |n',\mathbf{k}'_{1}\rangle$, and $|f\rangle \equiv |\ks\rangle \otimes |l, \mathbf{k}_{b} \rangle$.

It is convenient to transform Eq.~(\ref{eq:P_continuus}) from the $S$-matrix to $T$-matrix representation, connected to each other by the relation $S_{fi} \equiv \delta_{fi} - i 2\pi \delta(E_f-E_i)\mathscr{T}_{fi}$, which enforces the conservation of the total energy between the initial and the final states of the system. Furthermore, it is useful to enforce the conservation of the total linear momenta, such that $\mathcal{P}_i=\bk_0+\bk_1=\mathcal{P}_f=\ks+\bk_b$, and set $\mathscr{T}_{fi}=\delta(\mathcal{P}_f-\mathcal{P}_i) T_{fi}$. Combining all these expressions together and utilizing the conditions that $|i\rangle \neq |f\rangle$ and $|i'\rangle \neq |f\rangle$, the partial scattering probability can be written as
\begin{equation}
    \begin{split}
        P_l(\ks)=
        \int d\mathbf{k}_b
        \sum_{n'}\sum_{n}
        & C_{n'}^* C_n
        \int d \mathbf{k}_{0}^\prime 
        \int d \mathbf{k}_{0}
        \int d \mathbf{k}_{1}^\prime        
        \int d \mathbf{k}_{1} ~
        a^*_0(\mathbf{k}_{0}^\prime)
        a_0(\mathbf{k}_{0}) 
        a^*_1(\mathbf{k}_{1}^\prime)
        a_1(\mathbf{k}_{1}) 
        \\
        \times &
        (2 \pi)^2
        \delta(E_f-E_i)
        \delta(E_f-E_{i'})
        \delta(\mathcal{P}_f-\mathcal{P}_i)
        \delta(\mathcal{P}_f-\mathcal{P}_{i'})
        T^*_{fi^\prime}
        T_{fi},
    \end{split}
\end{equation}
\end{widetext}
which, upon using the relation $d^2 P_l / dE\tsb{s}d\Omega = |\ks| P_l (\ks)$ for the partial doubly differential electron scattering probability, exactly reproduces Eq.~(14) of Ref.~\cite{shao2014}. In that work, the corresponding expression was derived using the Lippmann--Schwinger formalism for the transition operator. While, under certain assumptions, this alternative derivation is fully consistent with the time-dependent approach introduced earlier in this work, we note that Eq.~(\ref{eq:General_Scattering_Prob}) is more general, as it allows for mixed states and enables the inclusion of internal interactions between particles in the beam, which could be important for proper description of realistic diffraction measurements. Indeed, interactions between photons in x-ray beams suitable for a typical diffraction experiment are negligible, such that the probe can be accurately described as a collection of noninteracting particles. In contrast, for electron beams, Coulomb interactions between electrons can play a significant role, influencing the beam's coherence properties and temporal and spatial structure, and should therefore be taken into account for a consistent description of electron diffraction experiments.

\section{X-ray scattering}
\subsection{Time-independent case}
\label{sec:Xray_st}
Considering elastic x-ray scattering from a stationary target, we first note that, due to the absence of direct photon-photon interactions at low field intensities, multiple photons can occupy the same momentum mode and therefore obey Bose--Einstein statistics. As a result, the initial state of a continuous x-ray beam can be described as $|\psi^{\ki}_i\rangle=|n_{\bk\tsb{in},\lambda\tsb{in}}\rangle$, where $n_{\bk\tsb{in},\lambda\tsb{in}}$ denotes the number of photons occupying the momentum mode $(\bk\tsb{in},\lambda\tsb{in})$, with $\lambda\tsb{in}$ denoting the polarization index. The corresponding scattered state $|\psi^{\ks}_f\rangle=|n_{\bk\tsb{in},\lambda\tsb{in}}-1,1_{\bk\tsb{s},\lambda\tsb{s}}\rangle$ is obtained by transitioning one photon from the initial mode to the momentum mode $(\bk\tsb{s},\lambda\tsb{s})$. Accordingly, the matrix element of the interaction Hamiltonian for the quantized x-ray beam can be written as:
\begin{widetext}
\begin{equation}
    \langle \psi^{\ks}_m | \hat{H}^{\text{x-ray}}_{\text{int}} | \psi^{\ki}_j \rangle = 
        \frac{2\pi}{V}
        \sqrt{\frac{1}{\omega_{\bk\tsb{in}} \omega_{\bk\tsb{s}}}}
        \left(  \be[\tsb{s}]^* \cdot \be[\tsb{in}] \right)
            \sqrt{n_{\bk\tsb{in},\lambda\tsb{in}}}
        \int d^3 r \hat{\rho}(\mathbf{r})
        e^{i (\bk\tsb{in} - \bk\tsb{s}) \cdot \mathbf{r}},
\end{equation}
%
where $\omega_{\bk\tsb{in}}$ and $\omega_{\bk\tsb{s}}$ are the photon energies of the incoming and the scattered x-ray beam, respectively, $\be[\tsb{in}]$ and $\be[\tsb{s}]$ are their polarization vectors, and $\hat{\rho}(\mathbf{r})$ is the electron density operator of the target. The detailed derivation of this expression is presented in Appendix~\ref{app:xray_beam}.

Substituting the matrix element $\langle \psi^{\ks}_m | \hat{H}^{\text{x-ray}}_{\text{int}} | \psi^{\ki}_j \rangle$ to Eq.~(\ref{eq:Pks_stationary}), we obtain the following expression for the scattering probability to find a photon in the mode $(\ks,\lambda\tsb{s})$ from the initial mode $(\ki,\lambda\tsb{in})$:
%
\begin{equation}
\label{eq:Pks_Xray_stationary}
	P(\ks,\lambda\tsb{s}) = \frac{(2\pi)^3}{V^2} T \delta(E\tsp{b}_f-E\tsp{b}_i)
		\frac{1}{\omega_{\bk\tsb{in}} \omega_{\bk\tsb{s}}}
		\left|  \be[\tsb{s}]^* \cdot \be[\tsb{in}] \right|^2
			n_{\bk\tsb{in},\lambda\tsb{in}}   
		\left |
		\int d^3 r \langle \Phi_i | 
			\hat{\rho}(\mathbf{r}) 
		| \Phi_i \rangle e^{i (\bk\tsb{in} - \bk\tsb{s}) \cdot \mathbf{r}}
		\right |^2.
\end{equation}
%
Substituting Eq.~(\ref{eq:Pks_Xray_stationary}) into Eq.~(\ref{eq:dP}) and assuming that the polarization of the scattered photon remains unobserved, such that the scattering probability to find a photon with momentum $\ks$ is obtained by performing the summation over the polarization index $\lambda\tsb{s}$, we obtain
%
\begin{equation}
\label{eq:dP_Xray_st}
    \frac{dP}{d\Omega} = \frac{\alpha^3}{V} T 
        \sum_{\lambda\tsb{s}} \left|  \be[\tsb{s}]^* \cdot \be[\tsb{in}] \right|^2
        \left |
		\int d^3 r \langle \Phi_i | 
			\hat{\rho}(\mathbf{r})
		| \Phi_i \rangle e^{i (\bk\tsb{in} - \bk\tsb{s}) \cdot \mathbf{r}}
	\right |^2,
\end{equation}
\end{widetext}
where the integration has been performed using the relations $|\bk|=\alpha \omega_{\bk}$ and $E\tsp{b}_f-E\tsp{b}_i=n_{\bk\tsb{in},\lambda\tsb{in}}(\omega_f-\omega_i)$, together with the scaling property of the delta function, which yields
\begin{equation}
    \delta(E\tsp{b}_f-E\tsp{b}_i)=\frac{1}{n_{\bk\tsb{in},\lambda\tsb{in}}}\delta(\omega_f-\omega_i).
\end{equation}

To obtain the differential scattering cross-section, we express the x-ray photon flux for a single-mode plane wave in a quantization volume $V$ with $n_{\bk\tsb{in},\lambda\tsb{in}}$ photons as
\begin{equation}
\label{eq:flux_Xray_st}
    F = \frac{1}{\alpha} \frac{n_{\bk\tsb{in},\lambda\tsb{in}}}{V},
\end{equation}
which follows from the fact that all photons propagate with the speed of light in a well-defined direction. Since the flux is time-independent, its integral over the duration of the experiment, i.e. the fluence, is trivial: $\int_{t_i}^{t_f} F dt = F T$. Substituting Eq.~(\ref{eq:dP_Xray_st}) and the obtained fluence into the analog of Eq.~(\ref{eq:sigma_P}) for the differential cross section, we obtain
\begin{equation}
\label{eq:dsigma_Xray_st}
    \frac{d\sigma}{d\Omega} = \frac{d\sigma\tsb{th}}{d\Omega}
        \left |
		\int d^3 r \langle \Phi_i | 
			\hat{\rho}(\mathbf{r})
		| \Phi_i \rangle e^{i (\bk\tsb{in} - \bk\tsb{s}) \cdot \mathbf{r}}
	\right |^2,
\end{equation}
where
\begin{equation}
    \frac{d\sigma\tsb{th}}{d\Omega} = \alpha^4 \sum_{\lambda\tsb{s}} \left| \be[\tsb{s}]^* \cdot \be[\tsb{in}] \right|^2
\end{equation}
is the differential Thomson scattering cross section (see Ref.~\cite{santra2009}). The obtained expression, Eq.~(\ref{eq:dsigma_Xray_st}), is the well-known result of scattering theory (see, e.g., Refs.~\cite{coppens1992,coppens1997}), showing that the differential x-ray scattering cross section is given by the modulus squared of the Fourier transform of the electron density of the target. The prefactor $d\sigma\tsb{th}/d\Omega$ sets the overall scale of the process and corresponds to the scattering of the electromagnetic field from a free electron.

\subsection{Time-dependent case}
\label{sec:Xray_td}
Given the negligible photon-photon interaction in x-ray beam, photon number states constitute natural eigenstates of the beam Hamiltonian. Consequently, the matrix elements $\langle \ns{m}| \hat{U}\tsb{b}(t_f,t_1) \hat{H}\tsp{x-ray}_{\text{int}} \hat{U}\tsb{b}(t_1,t_i) |\ns{n} \rangle$ and $\langle \ns{\bar{n}}| \hat{U}^{\dagger}\tsb{b}(t_2,t_i) \left( \hat{H}\tsp{x-ray}_{\text{int}} \right)^{\dagger} \hat{U}^{\dagger}\tsb{b}(t_f,t_2) |\ns{m} \rangle$ contributing to the general Eq.~(\ref{eq:General_Scattering_Prob}) can be written as
\begin{widetext}
\begin{equation}
\label{eq:U_xray_me}
\begin{split}
    \langle \ns{m}| & \hat{U}\tsb{b}(t_f,t_1) \hat{H}\tsp{x-ray}_{\text{int}} \hat{U}\tsb{b}(t_1,t_i) |\ns{n} \rangle 
    = e^{-i E_{\ns{m}}(t_f-t_1)} e^{-i E_{\ns{n}}(t_1-t_i)} \\
    & \times    \frac{2\pi}{V}
		\sqrt{\frac{1}{\omega_{\ki} \omega_{\ks}}}
			\left(  \bm{\varepsilon}_{\ki,\lambda\tsb{in}} \cdot \bm{\varepsilon}^*_{\ks,\lambda_s} \right)
            \sum_{\mathbf{k}_1,\lambda_1} \langle \ns{m} | \cop[\tsb{s}]\aop[_1] | \ns{n} \rangle
        \int d^3 r_1 \hat{\rho}(\mathbf{r}_1)
	e^{i (\bk_1 - \bk\tsb{s}) \cdot \mathbf{r}_1},
\end{split}
\end{equation}
and
\begin{equation}
\label{eq:U_xray_me_cc}
\begin{split}
    \langle \ns{\bar{n}}| & \hat{U}^{\dagger}\tsb{b}(t_2,t_i) \left( \hat{H}\tsp{x-ray}_{\text{int}} \right)^{\dagger} \hat{U}^{\dagger}\tsb{b}(t_f,t_2) |\ns{m} \rangle =
        e^{i E_{\ns{\bar{n}}}(t_2-t_i)} e^{i E_{\ns{m}}(t_f-t_2)} \\
        & \times \frac{2\pi}{V}
		\sqrt{\frac{1}{\omega_{\ki} \omega_{\ks}}}
			\left(  \bm{\varepsilon}^*_{\ki,\lambda\tsb{in}} \cdot \bm{\varepsilon}_{\ks,\lambda_s} \right)
            \sum_{\mathbf{k}_2,\lambda_2} \langle \ns{\bar{n}} | \cop[_2]\aop[\tsb{s}] | \ns{m} \rangle
        \int d^3 r_2 \hat{\rho}(\mathbf{r}_2)
	e^{-i (\bk_2 - \bk\tsb{s}) \cdot \mathbf{r}_2},
\end{split}
\end{equation}
%
where $E_{\ns{m}}$, $E_{\ns{n}}$, and $E_{\ns{\bar{n}}}$ are the eigenenergies of the corresponding beam states $|\ns{m}\rangle$, $|\ns{n}\rangle$, and $|\ns{\bar{n}}\rangle$, respectively, and $\cop$ and $\aop$ are the bosonic creation and annihilation field operators, respectively, responsible for adding and removing a photon in the single-particle mode with momentum $\bk$ and polarization index $\lambda$. The detailed derivation of Eqs.~(\ref{eq:U_xray_me}) and~(\ref{eq:U_xray_me_cc}), as well as the discussion of the properties of the corresponding field operators, is presented in Appendix~\ref{app:xray_beam}.

Substituting Eqs.~(\ref{eq:U_xray_me}) and~(\ref{eq:U_xray_me_cc}) to Eq.~(\ref{eq:General_Scattering_Prob}), we obtain the following expression for the scattering probability
%
\begin{equation}
\label{eq:Pksls_xray}
\begin{split}
    P(\ks,\lambda\tsb{s})=\lim_{\substack{t_i \to -\infty \\ t_f\to+\infty}}  
        & \sum_{\ns{m}} \sum_{\ns{n},\ns{\bar{n}}}   
        \rho\tsp{b}_{\ns{n},\ns{\bar{n}}}  
        \int_{t_i}^{t_f} dt_1 \int_{t_i}^{t_f} dt_2 \
        e^{-i E_{\ns{\bar{n}}} t_i}
        e^{ i E_{\ns{n}} t_i}
        e^{ i t_2 (E_{\ns{\bar{n}}} - E_{\ns{m}})}
        e^{-i t_1 (E_{\ns{n}} - E_{\ns{m}})} \\
    \times & 
        \left( \frac{2\pi}{V} \right)^2 \frac{1}{\omega_{\ki} \omega_{\ks}}
        \left|  \be[\tsb{s}]^* \cdot \be[\tsb{in}] \right|^2
        \sum_{\mathbf{k}_1,\lambda_1} \sum_{\mathbf{k}_2,\lambda_2}
            \langle \ns{\bar{n}} | \cop[_2]\aop[\tsb{s}] | \ns{m} \rangle
            \langle \ns{m} | \cop[\tsb{s}]\aop[_1] | \ns{n} \rangle \\
    \times &
        \int d^3 r_1 \int d^3 r_2 \
        e^{-i \bk\tsb{s} \cdot (\mathbf{r}_1-\mathbf{r}_2)} 
        e^{ i \bk_1 \cdot \mathbf{r}_1} e^{-i \bk_2 \cdot \mathbf{r}_2} 
        \langle \Phi(t_2)|
            \hat{\rho}(\mathbf{r}_2)
            \hat{U}\tsb{t}(t_2,t_1)
            \hat{\rho}(\mathbf{r}_1) 
        |\Phi(t_1)\rangle.
\end{split}
\end{equation}
Rearranging terms in Eq.~(\ref{eq:Pksls_xray}) and performing the summation over the polarization index $\lambda\tsb{s}$, we can write
\begin{equation}
\label{eq:Pks_xray}
\begin{split}
    P(\ks)=\lim_{\substack{t_i \to -\infty \\ t_f\to+\infty}}  
    &  
        \frac{2\pi}{V} \frac{1}{(\omega_{\mathbf{k}\tsb{in}})^2 \omega_{\ks}}
        \frac{1}{\alpha^4}\frac{d \sigma\tsb{th}}{d\Omega}
        \int_{t_i}^{t_f} dt_1 
        \int_{t_i}^{t_f} dt_2 \
            e^{-i \omega_{\ks}(t_2 - t_1)} \\
    \times & 
        \int d^3 r_2
        \int d^3 r_1 \
        e^{-i \bk\tsb{s} \cdot (\mathbf{r}_1 - \mathbf{r}_2)}         
        \langle \Phi(t_2)|
            \hat{\rho}(\mathbf{r}_2) 
            \hat{U}\tsb{t}(t_2,t_1)
            \hat{\rho}(\mathbf{r}_1)
        |\Phi(t_1)\rangle
        G\tsp{(1)}\tsb{x-ray}(\br[\tsb{2}],t_2,\br[\tsb{1}],t_1),
\end{split}
\end{equation}
where most of the terms responsible for the properties of the beam have been collected under the function $G\tsp{(1)}\tsb{x-ray}(\br[\tsb{2}],t_2,\br[\tsb{1}],t_1)$:
\begin{equation}
\label{eq:G_xray}
\begin{split}
    G\tsp{(1)}\tsb{x-ray}(\br[\tsb{2}],t_2,\br[\tsb{1}],t_1) &
    = \frac{2\pi \omega_{\bk_{\tsb{in}}}}{V}
        \sum_{\mathbf{k}_1,\lambda_1} \sum_{\mathbf{k}_2,\lambda_2}
            e^{-i \omega_{\bk\tsb{1}} t_1}
            e^{ i \omega_{\bk\tsb{2}} t_2}
            e^{-i \bk\tsb{2} \cdot \mathbf{r}_2} 
            e^{ i \bk\tsb{1} \cdot \mathbf{r}_1} \\
        & \times \sum_{\ns{m}} \sum_{\ns{n},\ns{\bar{n}}}   
            \rho\tsp{b}_{\ns{n},\ns{\bar{n}}}
                e^{-i E_{\ns{\bar{n}}} t_i}
                e^{ i E_{\ns{n}} t_i}
            \langle \ns{\bar{n}} | \cop[_2]
                \aop[\tsb{s}] | \ns{m} \rangle \langle \ns{m} | \cop[\tsb{s}]
            \aop[_1] | \ns{n} \rangle \\
    = & \frac{2\pi \omega_{\bk_{\tsb{in}}}}{V}
    \sum_{\mathbf{k}_1,\lambda_1} \sum_{\mathbf{k}_2,\lambda_2}
    \operatorname{Tr}\left[ \hat{\tilde{\rho}}\tsb{b}
    \cop[\tsb{2}]  \aop[\tsb{1}]\right]
    e^{-i \omega_{\bk\tsb{1}} t_1}
    e^{ i \omega_{\bk\tsb{2}} t_2}
    e^{-i \bk\tsb{2} \cdot \mathbf{r}_2} 
    e^{ i \bk\tsb{1} \cdot \mathbf{r}_1}.
\end{split}
\end{equation}
\end{widetext}
To obtain the above expressions, we utilized the fact that the energy differences between the initial and the scattered states of the beam can be expressed as $E_{\ns{\bar{n}}} - E_{\ns{m}} = \omega_{\bk_2} - \omega_{\ks}$ and $E_{\ns{n}} - E_{\ns{m}} = \omega_{\bk_1} - \omega_{\ks}$, given the fact that the corresponding states are different from each other by transition of only one photon between the corresponding field modes. Furthermore, we absorbed the time-dependent phase factors $e^{-i E_{\ns{\bar{n}}} t_i}$ and $e^{ i E_{\ns{n}} t_i}$, together with the corresponding density matrix elements $\rho\tsp{b}_{\ns{n},\ns{\bar{n}}}$, into a new density matrix $\hat{\tilde{\rho}}\tsb{b}$, which characterizes the initial state of the beam. Finally, we utilized the fact that annihilating a $(\ks,\lambda\tsb{s})$ photon from the number states $|\ns{m}\rangle$ generates a complete basis for representing photon states with one fewer particle, which makes it possible to take out the following resolution of identity: $\sum_{\ns{m}} \aop[\tsb{s}] | \ns{m} \rangle \langle \ns{m} | \cop[\tsb{s}] = \mathbb{I}$.

The expression in Eq.~(\ref{eq:G_xray}) can be identified as the approximate first-order radiation field correlation function~\cite{glauber1963,glauber1963a}:
\begin{equation}
    G\tsp{(1)}\tsb{x-ray}(\br[\tsb{2}],t_2,\br[\tsb{1}],t_1) 
    \approx \langle \hat{E}^{(-)}(\mathbf{r}_2,t_2) \hat{E}^{(+)}(\mathbf{r}_1,t_1) \rangle\tsb{b},
\end{equation}
where the positive and negative frequency electric field operators are defined as
\begin{equation}
    \hat{E}^{(+)}(\mathbf{r},t)=i\sum_{\bk,\lambda} \sqrt{\frac{2\pi \omega_{\bk}}{V}} \aop e^{i(\bk \cdot \mathbf{r} - \omega_{\bk}t)}
\end{equation}
and
\begin{equation}
    \hat{E}^{(-)}(\mathbf{r},t)=- i\sum_{\bk,\lambda} \sqrt{\frac{2\pi \omega_{\bk}}{V}} \cop e^{-i(\bk \cdot \mathbf{r} - \omega_{\bk}t)},
\end{equation}
respectively. The correlation function $G\tsp{(1)}\tsb{x-ray}(\br[\tsb{2}],t_2,\br[\tsb{1}],t_1)$ characterizes the spatial and temporal coherence properties of the radiation field. For a coherent ensemble of pulses it reduces to~\cite{dixit2012}
\begin{equation}
\begin{split}
    G\tsp{(1)}\tsb{x-ray}&(\br[\tsb{2}],t_2,\br[\tsb{1}],t_1) 
    \\ \approx & \ 2 \pi \alpha \omega_{\ki} I(\mathbf{r}_0, \gamma) C(\delta) e^{i \omega_{\mathbf{k}_{\mathrm{in}}} \delta} e^{i \mathbf{k}_{\mathrm{in}} \cdot\left(\mathbf{r}_1-\mathbf{r}_2\right)} 
\end{split}
\end{equation}
where $I(\mathbf{r}_0, \gamma)$ is the pulse intensity, $C(\delta)$ is a function of the pulse duration, $\mathbf{r}_0$ is the position of the target, $\gamma = \frac{t_2+t_1}{2}$, and $\delta = t_2-t_1$.


\section{Non-relativistic electron scattering}
\subsection{Time-independent case}
\label{sec:el_st}

In contrast to the x-ray case considered in Sec.~\ref{sec:Xray_st}, the presence of self-interactions between electrons in the beam, together with their fermionic nature governed by Fermi--Dirac statistics, prohibits the preparation of a continuous beam in which multiple electrons occupy the same $(\ki,\sigma)$ mode. Therefore, the initial and the final states of the beam, $|\psi^{\ki}_i\rangle=|1_{\bk\tsb{in},\sigma}\rangle$ and $|\psi^{\ks}_f\rangle=|1_{\bk\tsb{s},\sigma}\rangle$, respectively, contain only a single electron transitioning from the initial mode $(\bk\tsb{in},\sigma)$ to the scattered mode $(\bk\tsb{s},\sigma)$. Note that only transitions that preserve the electronic spin $\sigma$ are allowed, so that the overall process of electron scattering is independent of the spin. The corresponding matrix element of the interaction Hamiltonian between the electron beam and the target can be written as
\begin{equation}
     \langle \psi^{\ks}_m | \hat{H}^{\text{el.}}_{\text{int}} | \psi^{\ki}_j \rangle = 
        \frac{4\pi}{V} 
            \frac{1}{|\ki-\ks|^2}
            \int d^3 r \hat{\rho}(\mathbf{r}) e^{i (\bk\tsb{in} - \bk\tsb{s}) \cdot \mathbf{r}}.
\end{equation}
While the resulting expression is well known (see, e.g., Ref.~\cite{centurion2022}), its derivation within the quantum field theory framework, presented in Appendix~\ref{app:el_beam}, makes the direct correspondence between x-ray and electron scattering explicit.

Following a similar logic to the one discussed in Sec.~\ref{sec:Xray_st}, i.e. substituting the matrix element $\langle \psi^{\ks}_m | \hat{H}^{\text{el.}}_{\text{int}} | \psi^{\ki}_j \rangle$ to Eq.~(\ref{eq:Pks_stationary}), we obtain the following expression for the scattering probability to find an electron in the mode $(\ks,\sigma)$ from the initial mode $(\ki,\sigma)$:
\begin{widetext}
\begin{equation}
\label{eq:Pks_el_stationary}
	P(\ks) = \frac{4(2\pi)^3}{V^2} T \delta(E\tsp{b}_f-E\tsp{b}_i)
		\frac{1}{|\ki-\ks|^4}
		\left |
		\int d^3 r \langle \Phi_i | 
			\hat{\rho}(\mathbf{r}) 
		| \Phi_i \rangle e^{i (\bk\tsb{in} - \bk\tsb{s}) \cdot \mathbf{r}}
		\right |^2.
\end{equation}
\end{widetext}

Substituting Eq.~(\ref{eq:Pks_el_stationary}) to the general Eq.~(\ref{eq:dP}) for the differential scattering probability and using the relations $|\bk|=\sqrt{2E}$ and $d|\bk|=\frac{1}{\sqrt{2E}}dE=\frac{1}{|\bk|}dE$ to carry out the integration, we obtain
\begin{equation}
\label{eq:dP_el_st}
    \frac{dP}{d\Omega} = \frac{4}{V} T 
        \frac{|\ks|}{|\ki - \ks|^4}
        \left |
		\int d^3 r \langle \Phi_i | 
			\hat{\rho}(\mathbf{r}) 
		| \Phi_i \rangle e^{i (\ks - \ki) \cdot \mathbf{r}}
	\right |^2.
\end{equation}
%

To finish the derivation and arrive to the expression for the differential scattering cross-section, we note that the flux of an electron present in the plane-wave state $\psi^{\bk}_j(\mathbf{r},t)=1/\sqrt{V}e^{i(\bk \cdot \mathbf{r}-\omega t)}$ along the propagation direction is $F=|\bk|/V$. Substituting the corresponding fluence, $\int_{t_i}^{t_f} F dt = F T$, and Eq.~(\ref{eq:dP_el_st}) into the analog of Eq.~(\ref{eq:sigma_P}) for the differential cross section, we obtain
\begin{equation}
\label{eq:dsigma_el_st}
    \frac{d\sigma}{d\Omega} =
        \frac{d\sigma_{\text{ruth}}}{d\Omega}
        \left |
		\int d^3 r \langle \Phi_i | 
			\hat{\rho}(\mathbf{r})
		| \Phi_i \rangle e^{i (\bk\tsb{in} - \bk\tsb{s}) \cdot \mathbf{r}}
	\right |^2.
\end{equation}
Here, 
\begin{equation}
    \frac{d\sigma_{\text{ruth}}}{d\Omega} = \frac{4}{|\ki - \ks|^4} = \frac{1}{16 E^2 \sin^4{\theta/2}}
\end{equation}
is the differential Rutherford scattering cross section, and we utilized the relations $|\ki|=|\ks|=|\bk|=\sqrt{2E}$ and $|\ki - \ks|=2|\bk| \sin{\frac{\theta}{2}}$, with $\theta$ being the angle between the initial incoming and the final outgoing momenta of the scattered particle, to arrive to the familiar textbook formula. Similarly to Eq.~(\ref{eq:dsigma_Xray_st}) for the x-ray scattering cross section, the obtained Eq.~(\ref{eq:dsigma_el_st}) for the electron scattering shows that the cross section is given by the modulus squared of the Fourier transform of the electron density of the target. The prefactor $d\sigma_{\text{ruth}}/d\Omega$ sets the overall scale of the process and corresponds to the deflection of a single electron from a long-range Coulomb potential.

\subsection{Time-dependent case}
\label{sec:td_el_nonrel}

The presence of self-interactions within the electron beam complicates the description of time-dependent scattering processes. Since the electron number states are not the eigenstates of the beam Hamiltonian, the time evolution present in the matrix elements $\langle \ns{m}| \hat{U}\tsb{b}(t_f,t_1) \hat{H}\tsp{el.}_{\text{int}} \hat{U}\tsb{b}(t_1,t_i) |\ns{n} \rangle$ and $\langle \ns{\bar{n}}| \hat{U}^{\dagger}\tsb{b}(t_2,t_i) \left( \hat{H}\tsp{el.}_{\text{int}} \right)^{\dagger} \hat{U}^{\dagger}\tsb{b}(t_f,t_2) |\ns{m} \rangle$ must be, at least partially, retained. Nevertheless, after appropriate transformations (see Appendix~\ref{app:el_beam} for the detailed derivation) these matrix elements can be written as:
\begin{widetext}
\begin{equation}
\label{eq:U_el_me}
\begin{split}
    \langle\{m\}| 
    & \hat{U}_{\mathrm{b}}\left(t_f, t_1\right) 
    \hat{H}\tsp{el.}_{\mathrm{int}} 
    \hat{U}_{\mathrm{b}}\left(t_1, t_i\right)
    |\{n\}\rangle
    =  \, e^{-i \varepsilon_{\ks}(t_f-t_1)} \\
    & \times \frac{2\pi}{V}  
    \frac{2}{|\bk\tsb{in}-\ks|^2}     
    \sum_{\bk_1,\sigma_1}
    \langle\{m\}|
        \hat{\mathrm{b}}^{\dagger}_{\ks,\sigma\tsb{s}} 
        \hat{U}_{\mathrm{b}}\left(t_f, t_1\right)
        \hat{\mathrm{b}}_{\bk_1,\sigma_1}
        \hat{U}_{\mathrm{b}}\left(t_1, t_i\right)
    |\{n\}\rangle 
    \int d^3 r_1 ~\hat{\rho}(\mathbf{r}_1)    
        e^{i (\bk_1 - \ks) \cdot \mathbf{r}_1}
\end{split}
\end{equation}
and
\begin{equation}
\label{eq:U_el_me_cc}
\begin{split}
    \langle \ns{\bar{n}}|
    &   \hat{U}_{\mathrm{b}}^\dagger\left(t_2, t_i\right)  
        \left( \hat{H}\tsp{el.}_{\text{int}} \right)^{\dagger} 
        \hat{U}_{\mathrm{b}}^\dagger\left(t_f, t_2\right)
    |\ns{m} \rangle
    = \, e^{i \varepsilon_{\ks}(t_f-t_2)} \\
    & \times \frac{2\pi}{V}
    \frac{2}{|\ks - \bk\tsb{in}|^2}         
    \sum_{\bk_2,\sigma_2}
    \langle\{\bar{n}\}| 
        \hat{U}_{\mathrm{b}}^\dagger\left(t_2, t_i\right) 
        \hat{\mathrm{b}}^{\dagger}_{\bk_2,\sigma_2}
        \hat{U}_{\mathrm{b}}^\dagger\left(t_f, t_2\right)
        \hat{\mathrm{b}}_{\ks,\sigma\tsb{s}} 
    |\ns{m}\rangle
    \int d^3 r_2 ~
        \hat{\rho}(\mathbf{r}_2)        
        e^{-i (\bk_2 - \ks)  \cdot \mathbf{r}_2},
\end{split}
\end{equation}
where $\varepsilon_{\ks}$ denotes the energy of the beam electron with momentum $\ks$, and $\hat{\mathrm{b}}^{\dagger}_{\bk,\sigma}$ and $\hat{\mathrm{b}}_{\bk,\sigma}$ are the fermionic creation and annihilation operators, respectively, responsible for adding and removing an electron in the single-particle mode with momentum $\bk$ and spin $\sigma$.

Substituting Eqs.~(\ref{eq:U_el_me}) and (\ref{eq:U_el_me_cc}) to Eq.~(\ref{eq:General_Scattering_Prob}) and rearranging the terms, we obtain the following expression for the time-dependent electron scattering probability:
\begin{equation}
\label{eq:Pks_nrel}
\begin{split}
    P(\ks)=\lim_{\substack{t_i \to -\infty \\ t_f\to+\infty}}  
    &  
        \frac{(2\pi)^2}{V} \frac{d \sigma\tsb{ruth}}{d\Omega}
        \int_{t_i}^{t_f} dt_1 
        \int_{t_i}^{t_f} dt_2 \
            e^{-i \varepsilon_{\ks}(t_2 - t_1)} \\
    \times & 
        \int d^3 r_2
        \int d^3 r_1 \
        e^{-i \bk\tsb{s} \cdot (\mathbf{r}_1 - \mathbf{r}_2)}         
        \langle \Phi(t_2)|
            \hat{\rho}(\mathbf{r}_2) 
            \hat{U}\tsb{t}(t_2,t_1)
            \hat{\rho}(\mathbf{r}_1)
        |\Phi(t_1)\rangle
        G\tsp{(1)}\tsb{el.}(\br[\tsb{2}],t_2,\br[\tsb{1}],t_1).
\end{split}
\end{equation}
Similarly to the x-ray case considered in Sec.~\ref{sec:Xray_td}, we collected most of the terms responsible for the properties of the beam under the function $G\tsp{(1)}\tsb{el.}(\br[\tsb{2}],t_2,\br[\tsb{1}],t_1)$:
\begin{equation}
\label{eq:Fermionic_1st_correlation}
\begin{split}
    G\tsp{(1)}\tsb{el.}( & \br[\tsb{2}],t_2,\br[\tsb{1}],t_1)
    = \frac{1}{V}
        \sum_{\mathbf{k}_1,\sigma_1} \sum_{\mathbf{k}_2,\sigma_2}
            e^{-i \bk\tsb{2} \cdot \mathbf{r}_2} 
            e^{ i \bk\tsb{1} \cdot \mathbf{r}_1} \\
        \times & \sum_{\ns{m}} \sum_{\ns{n},\ns{\bar{n}}}   
            \rho\tsp{b}_{\ns{n},\ns{\bar{n}}}
    \langle\{\bar{n}\}| 
        \hat{U}_{\mathrm{b}}^\dagger\left(t_2, t_i\right) 
        \hat{\mathrm{b}}^{\dagger}_{\bk_2,\sigma_2}
        \hat{U}_{\mathrm{b}}^\dagger\left(t_f, t_2\right)
        \hat{\mathrm{b}}_{\ks,\sigma\tsb{s}} 
    |\ns{m}\rangle
    \langle\{m\}|
        \hat{\mathrm{b}}^{\dagger}_{\ks,\sigma\tsb{s}} 
        \hat{U}_{\mathrm{b}}\left(t_f, t_1\right)
        \hat{\mathrm{b}}_{\bk_1,\sigma_1}
        \hat{U}_{\mathrm{b}}\left(t_1, t_i\right)
    |\{n\}\rangle \\
    = & \sum_{\sigma_1,\sigma_2} \sum_{\ns{n},\ns{\bar{n}}}   
            \rho\tsp{b}_{\ns{n},\ns{\bar{n}}} 
        \langle\{\bar{n}\}| 
            \hat{U}^\dagger\tsb{b}\left(t_2, t_i\right)
            \hat{\psi}\tsb{b}^\dagger(\mathbf{r}_2,\sigma_2)
            \hat{U}\tsb{b}\left(t_2, t_1\right)
            \hat{\psi}\tsb{b}(\mathbf{r}_1,\sigma_1)
            \hat{U}\tsb{b}\left(t_1, t_i\right)
        |\{n\}\rangle \\
    = & \sum_{\sigma_1,\sigma_2}
    \langle 
        \hat{\psi}\tsb{b}^\dagger(\mathbf{r}_2,\sigma_2;t_2) 
        \hat{\psi}\tsb{b}(\mathbf{r}_1,\sigma_1;t_1)
    \rangle\tsb{b}
    = \sum_{\sigma}
    \langle 
        \hat{\psi}\tsb{b}^\dagger(\mathbf{r}_2,\sigma;t_2) 
        \hat{\psi}\tsb{b}(\mathbf{r}_1,\sigma;t_1)
    \rangle\tsb{b},
\end{split}
\end{equation}
\end{widetext}
which is the first-order electronic correlation function for a fermionic field. In the final step of the derivation, the anticommutation relations of the fermionic field operators were used, leading to the cancellation of the terms involving spin exchange.

\section{Relativistic electron scattering}
\label{sec:relel}
To incorporate relativistic corrections into electron-electron interactions, in Appendix~\ref{app:rel_beam} we developed an effective Hamiltonian that includes magnetic density-current and current-current couplings. For simplicity, we work in the small-angle, small-momentum-transfer regime relevant to the diffraction geometries considered below, and we neglect various spin-spin, spin-orbit, and retardation effects, since these contributions are expected to play a sub-leading role in the scattering signal from a time-dependent target. Indeed, the electron motion in the target can generate time-dependent currents, which in turn give rise to magnetic fields that couple directly to the incident electrons of the beam. While this contribution enter the Hamiltonian at a higher relativistic order than Coulomb interaction considered before, the time-dependent portion of the scattering signal coming from the relativistic density-current and current-current interactions could be comparable or even stronger than the one arising from the corresponding density-density coupling.

The matrix elements $\langle \ns{m}| \hat{U}\tsb{b}(t_f,t_1) \hat{H}\tsp{B}_{\text{int}} \hat{U}\tsb{b}(t_1,t_i) |\ns{n} \rangle$ and $\langle \ns{\bar{n}}| \hat{U}^{\dagger}\tsb{b}(t_2,t_i) \left(\hat{H}\tsp{B}_{\text{int}}\right)^{\dagger} \hat{U}^{\dagger}\tsb{b}(t_f,t_2) |\ns{m} \rangle$ for the relativistic Breit interaction Hamiltonian $\hat{H}\tsp{B}_{\mathrm{int}}$ can be written in natural units as (see Appendix~\ref{app:rel_beam} for the detailed derivation)
\begin{widetext}
\begin{equation}
\label{eq:U_rel_me}
\begin{split}
    \langle\{m\}| 
        & \hat{U}_{\mathrm{b}}\left(t_f, t_1\right) 
        \hat{H}\tsp{B}_{\mathrm{int}} 
        \hat{U}_{\mathrm{b}}\left(t_1, t_i\right)
    |\{n\}\rangle = e^{-i E_{\ks}(t_f-t_1)}
    \frac{\bar{e}^2}{2V}
    \frac{\widetilde D_{\mu_1\nu_1}(\ki - \ks)}{|\ki - \ks|^2} \frac{1}{\sqrt{E_{\ki} E_{\ks}}} \\
    & \times \sum_{\sigma_1}
    \bar{u}(k\tsb{s},\sigma\tsb{s})
    \gamma^{\nu_1} 
    u(k\tsb{in},\sigma_1) \sum_{\mathbf{k}_1}
    \langle\{m\}|
        \hat{\mathrm{b}}_{\ks,\sigma\tsb{s}} ^\dagger 
        \hat{U}_{\mathrm{b}}\left(t_f, t_1\right)  
        \hat{\mathrm{b}}_{\mathbf{k}_1,\sigma_1}
        \hat{U}_{\mathrm{b}}\left(t_1, t_i\right)
    |\{n\}\rangle
    \int d^3 r_1 ~
        \hat{J}_{\text{t}}^{\mu_1}(\mathbf{r}_1) e^{i(\bk_1 - \ks)\cdot\mathbf{r}_1},
\end{split}
\end{equation}
and
\begin{equation}
\label{eq:U_rel_me_cc}
    \begin{split}
        \langle \ns{\bar{n}}|
        &   \hat{U}_{\mathrm{b}}^\dagger\left(t_2, t_i\right)  
            \left( \hat{H}\tsp{B}_{\text{int}} \right)^{\dagger} 
            \hat{U}_{\mathrm{b}}^\dagger\left(t_f, t_2\right)
        |\ns{m} \rangle = e^{i E_{\ks}(t_f-t_2)} 
    \frac{\bar{e}^2}{2V}
    \frac{\widetilde D_{\mu_2\nu_2}(\ki - \ks)}{|\ki - \ks|^2} \frac{1}{\sqrt{E_{\ki} E_{\ks}}} \\
    & \times \sum_{\sigma_2}
    \bar{u}(k\tsb{in},\sigma_2)
    \gamma^{\nu_2} 
    u(k\tsb{s},\sigma\tsb{s})
    \sum_{\mathbf{k}_2}
        \langle\{\bar{n}\}| 
            \hat{U}_{\mathrm{b}}^\dagger\left(t_2, t_i\right) 
            \hat{\mathrm{b}}^{\dagger}_{\bk_2,\sigma_2}
            \hat{U}_{\mathrm{b}}^\dagger\left(t_f, t_2\right)
            \hat{\mathrm{b}}_{\ks,\sigma\tsb{s}} 
        |\ns{m}\rangle
        \int d^3 r_2 ~
            \hat{J}_{\text{t}}^{\mu_2}(\mathbf{r}_2)
            e^{-i (\bk_2 - \ks)  \cdot \mathbf{r}_2}.
    \end{split}
\end{equation}
Here, $\bar{e}^2$ is the electron charge, $E_{\bk} = \sqrt{\bk^2 + m^2}$ is the relativistic electron energy, $\bar{u}(k, \sigma)$ and $u(k, \sigma)$ are the Dirac spinors for an electron in $(k,\sigma)$ mode with $k=(E_{\mathbf{k}},\mathbf{k})$ denoting the four-momentum vector, $\gamma^{\nu}$ are the Dirac gamma matrices, and $\hat{J}_{\text{t}}^{\mu}(\mathbf{r})=(\hat{\rho}(\br), \hat{\mathbf{j}}(\br))$ is the Dirac four-current operator, encoding the charge density $\hat{\rho}(\br)$ and probability current $\hat{\mathbf{j}}(\br)$ of the target electron. The tensor $\widetilde D_{\mu\nu}(\ki - \ks)$ is defined as following
\begin{equation}
    \widetilde D_{\mu\nu}(\mathbf{S}) =
    \begin{cases}
        1 , & \mu=\nu=0,\\
        \frac{S_{\mu} S_{\nu}}{|\mathbf S|^2} - \delta_{\mu\nu},
        & \mu>0,\;\nu>0,\\
        0, & \text{otherwise},
    \end{cases}
\end{equation}    
where $S_{\mu}$ and $S_{\nu}$ are the elements of the four-momentum vector $S=(E_{\mathbf{S}},\mathbf{S})$.

Substituting Eqs.~(\ref{eq:U_rel_me}) and~(\ref{eq:U_rel_me_cc}) into Eq.~(\ref{eq:General_Scattering_Prob}), changing from natural to atomic units~\cite{baylis2006}, such that $\bar{e}^2=4\pi\alpha$, $\br\tsb{a.u.}=\alpha m_e \br\tsb{nat}$, $\bk\tsb{a.u.}=\bk\tsb{nat}/\alpha m_e$, $t\tsb{a.u.}=\alpha^2 m_e t\tsb{nat}$, $E\tsb{a.u.}=E\tsb{nat}/\alpha^2 m_e$, $\hat{\rho}\tsb{a.u.}=\hat{\rho}\tsb{nat}/(\alpha m_e)^3$, $\hat{\mathbf{j}}\tsb{a.u.}=\hat{\mathbf{j}}\tsb{nat}/\alpha(\alpha m_e)^3$, and rearranging the terms, we obtain the following expression for the time-dependent relativistic electron scattering probability:
\begin{equation}
\label{eq:Pks_rel}
\begin{split}
    P(\ks)=\lim_{\substack{t_i \to -\infty \\ t_f\to+\infty}}  
    &  
        \frac{(2\pi)^2}{V} \frac{d \sigma\tsb{ruth}}{d\Omega}
        \widetilde D_{\mu_1\nu_1}(\ki - \ks) \widetilde D_{\mu_2\nu_2}(\ki - \ks)
        \frac{k^{\nu_1}\tsb{in} k^{\nu_2}\tsb{in}}{E_{\ki} E_{\ks}}
        \int_{t_i}^{t_f} dt_1 
        \int_{t_i}^{t_f} dt_2 \
            e^{-i E_{\ks}(t_2 - t_1)} \\
    \times & 
        \int d^3 r_2
        \int d^3 r_1 \
        e^{-i \bk\tsb{s} \cdot (\mathbf{r}_1 - \mathbf{r}_2)}         
        \langle \Phi(t_2)|
            \hat{J}^{\mu_2}\tsb{t} (\mathbf{r}_2) 
            \hat{U}\tsb{t}(t_2,t_1)
            \hat{J}^{\mu_1}\tsb{t} (\mathbf{r}_1)
        |\Phi(t_1)\rangle
        G\tsp{(1)}\tsb{el.}(\br[\tsb{2}],t_2,\br[\tsb{1}],t_1),
\end{split}
\end{equation}
\end{widetext}
where $G\tsp{(1)}\tsb{el.}(\br[\tsb{2}],t_2,\br[\tsb{1}],t_1)$ is the first-order electronic correlation function for fermionic field defined earlier in Eq.~(\ref{eq:Fermionic_1st_correlation}). To derive the expression Eq.~(\ref{eq:Pks_rel}), we assumed that the spin of the scattered electrons remains unobserved, i.e. we summed over the spin index $\sigma\tsb{s}$, and performed contractions of spinors such that $\sum_{\sigma\tsb{s}} \bar{u}(k\tsb{in},\sigma\tsb{2}) \gamma^{\nu_1} u(k\tsb{s},\sigma\tsb{s}) \bar{u}(k\tsb{s},\sigma\tsb{s}) \gamma^{\nu_2} u(k\tsb{in},\sigma\tsb{1}) \approx 4k\tsb{in}^{\nu_1} k\tsb{in}^{\nu_2} \delta_{\sigma\tsb{2}\sigma\tsb{1}}$. The detailed discussion of the utilized contraction procedure is presented in Appendix~\ref{app:contraction}.

The obtained expression for the relativistic electron scattering probability, Eq.~(\ref{eq:Pks_rel}), incorporates density-density, density-current, and current-current interactions over all spatial degrees of freedom, while fully accounting for the orientation and energy of the incoming electron beam. The density-density part, as expected, is fully consistent with the non-relativstic formula, Eq.~(\ref{eq:Pks_nrel}), derived in Sec.~\ref{sec:td_el_nonrel}. The prefactor $k^{\nu_1}\tsb{in} k^{\nu_2}\tsb{in} / E_{\ki} E_{\ks}$ in Eq.~(\ref{eq:Pks_rel}) is just unity for the density-density contribution, since $k^0=E_{\bk}$, while it scales as $\alpha$ and $\alpha^2$, respectively, for the relativistic density-current and current-current interactions, which generally suppress their contributions to the total scattering probability. Since the beam is assumed to be non-relativistic in nature and only its interaction with the target is treated within an effective relativistic framework, its propagation in space and time is described by a simple first-order correlation function. As a result, the derived expression Eq.~(\ref{eq:Pks_rel}) is both practically convenient for applications and retains a clear physical interpretation in terms of charge and current correlations.

\section{Discussion}
\label{sec:discuss}

The presented derivations reveal a close correspondence between the scattering of x-rays and electrons across different regimes. For elastic scattering from a stationary target, the expressions for the differential x-ray and electron scattering cross sections, Eqs.~(\ref{eq:dsigma_Xray_st}) and~(\ref{eq:dsigma_el_st}), differ only by a prefactor that accounts for the scattering of the respective probe beam from a free electron~\cite{ma2020}. In both cases, the target properties are encoded in the squared absolute value of the Fourier transform of its electron density. Therefore, although the interaction mechanisms differ substantially in the two considered scenarios---x-rays couple to the target via electromagnetic forces, whereas electrons probe it through the Coulomb potential---the information about the target encoded in the diffraction signal is the same, being determined by the distribution of the electron density in the sample.

In the time-dependent case, the non-relativistic x-ray and electron scattering probabilities, Eqs.~(\ref{eq:Pks_xray}) and~(\ref{eq:Pks_nrel}), respectively, likewise preserve the close correspondence between the two probes, analogous to the time-independent scenario. Both expressions involve time-dependent correlation functions that encode the properties of both the target and the probe and enter the corresponding scattering formulas in an identical manner. In both cases, the dynamics of the target is captured by the density-density correlation function, governed by the temporal evolution of the electron density. The evolution of the beam, in turn, is contained in the corresponding field correlation function, which characterizes its temporal coherence and spectral distribution and determines how the probe propagates and interferes before and after the scattering process. In principle, probe-specific effects may arise in the time-dependent diffraction through the corresponding beam correlation functions thus reflecting distinct properties of x-ray and electron beams. Nevertheless, assuming a coherent probe prepared in a superposition of plane-wave modes, the propagation of both x-rays and electrons can be treated on an equal footing. Therefore, in appropriate limits, the diffraction of x-rays and electrons from a time-dependent target becomes virtually indistinguishable and can be described within a unified theoretical framework.

In fact, the developed formalism is sufficiently general to consistently and straightforwardly incorporate additional physical effects of interest. Considering relativistic interactions between the electron beam and the target, we derived the expression for the scattering probability, Eq.~(\ref{eq:Pks_rel}), that accounts for density-current and current-current coupling effects. Treating the required matrix elements within a relativistic quantum-field-consistent framework, the corresponding terms naturally emerge alongside the conventional non-relativistic density-density contribution. Therefore, Eq.~(\ref{eq:Pks_rel}) unifies the full set of leading interaction channels within a single expression, systematically combining charge and current responses of the dynamical target on equal footing.

In the following section, we employ Eq.~(\ref{eq:Pks_rel}) to illustrate the basic features of time-resolved scattering patterns and to analyze the respective roles of density-density, density-current, and current-current couplings in probing ultrafast laser-driven electron dynamics in neutral graphene sample.

\begin{center}
\begin{figure*}[t]
	\includegraphics[width=0.98\textwidth]{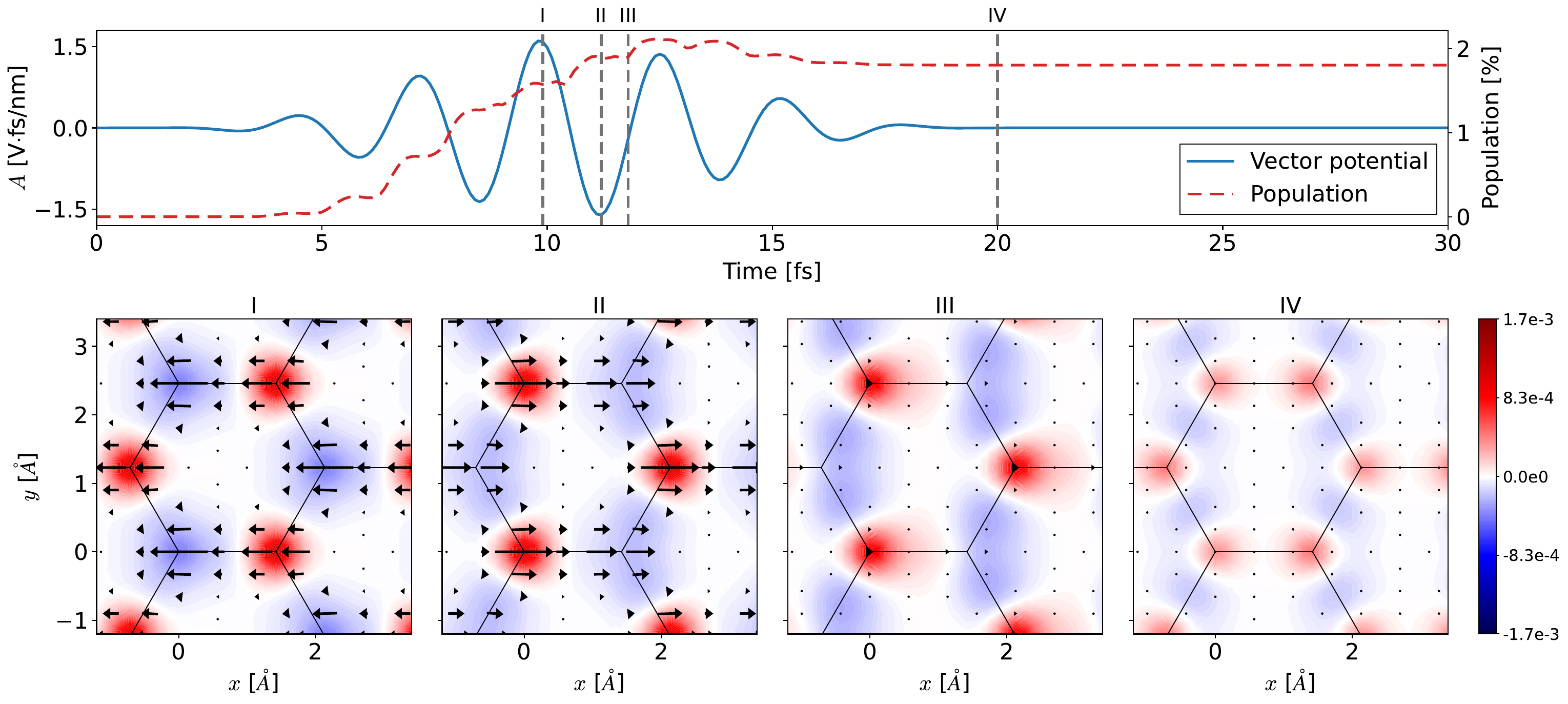}
\caption{Laser-driven electron dynamics in graphene. Top panel: The blue curve represents the vector potential of the pump laser field polarized along $x$ direction, while the red line depicts the evolution of the population of the conduction band of graphene. Bottom panel: The snapshots of the integrated real space electron difference density (contour plots) and the electron current density (arrows). The four columns depict the time instances $t_I=9.8$~fs, $t_{II}=11.2$~fs, $t_{III}=11.8$~fs, and $t_{IV}=20.0$~fs, indicated by the vertical grey dashed lines in top panel. More detailed discussion of the utilized computational procedure can be found in Ref.~\cite{yuan2025}.}
\label{fig:graphene_dynamics}
\end{figure*}
\end{center}

\vspace*{-1.5cm}

\section{Application to diffraction imaging of electron dynamics in solids}
\label{sec:applications}

To simulate the electron motion in graphene, we employ a fully quantum model based on numerical solution of the semiconductor Bloch equation combined with a tight-binding description of the electronic structure (see Ref.~\cite{yuan2025} for the details of the utilized procedure). We calculate the evolution of the electron density and electron current density in graphene driven by the interaction with a laser field of the following waveform: $\mathbf{E}(t)=\mathbf{e} E_0 \sin^4(\pi t/\tau)\cos(\omega t)$, where the peak field amplitude $E_0$ is chosen to be 2.5~V/nm, pulse duration $\tau$ is 21~fs, the photon energy $\omega$ is 1.55~eV which corresponds to the wavelength of 800~nm, and $\mathbf{e}$ denotes the unit vector in the direction of the field polarization. This choice of the target system, the utilized pump field parameters, and the consideration of electron diffraction is motivated by a recent experimental studies~\cite{hui2024,yuan2024a} (see also Ref.~\cite{baum2024_comment,hui2025_comment}) measuring ultrafast electron dynamics in graphene.

Figure~\ref{fig:graphene_dynamics} summarizes the results of the electron dynamics driven by the laser pulse described above and which is polarized along C--C bonds of the graphene sample ($x$ direction in our simulations). It is seen that the applied laser field (shown by blue solid line depicting the vector potential in top panel of Fig.~\ref{fig:graphene_dynamics}) causes the population transfer (shown by red dashed line) between the valence and conduction bands of graphene. The induced population transfer, together with the acceleration of the created electron-hole pairs by the applied laser field, triggers the ultrafast motion of the electron density (shown in bottom panel of Fig.~\ref{fig:graphene_dynamics}) and also creates the electron current (shown by arrows in bottom panel of Fig.~\ref{fig:graphene_dynamics}) in the graphene sample. The visualized electron density and the electron current density reflect the expectation values of the corresponding operators, $\hat{\rho}(\br)$ and $\hat{\mathbf{j}}(\br)$, respectively, which are the key ingredients present in Eq.~(\ref{eq:Pks_rel}).

Assuming that the temporal profile of the probe beam is infinitely short in time and that the scattering event takes place instantaneously, such that $t_1 = t_2$ in Eq.~(\ref{eq:Pks_rel}), the integral kernel of this equation can be written in application to the electron dynamics in solids as
\begin{widetext}
\begin{equation}
\begin{split}
    \int_{t_i}^{t_f} dt_1 
    \int_{t_i}^{t_f} dt_2 \
        e^{-i E_{\ks}(t_2 - t_1)} 
    & \int d^3 r_2
    \int d^3 r_1 \
    e^{-i \bk\tsb{s} \cdot (\mathbf{r}_1 - \mathbf{r}_2)}         
    \langle \Phi(t_2)|
        \hat{J}^{\mu_2}\tsb{t} (\mathbf{r}_2) 
        \hat{U}\tsb{t}(t_2,t_1)
        \hat{J}^{\mu_1}\tsb{t} (\mathbf{r}_1)
    |\Phi(t_1)\rangle
    G\tsp{(1)}\tsb{el.}(\br[\tsb{2}],t_2,\br[\tsb{1}],t_1) \\
    & \approx \frac{1}{\mathcal{N}} \sum_{n,m,f} \int_{\text{unit cell}} d \mathbf{p} ~ \rho_{n,m}(\mathbf{p},t)
			\mathcal{\hat{F}}^{*}_{\mathbf{S}}[J^{\mu_2}_{f,m} (\mathbf{p}_t,\mathbf{r})]
			\mathcal{\hat{F}}_{\mathbf{S}}[J^{\mu_1}_{f,n} (\mathbf{p}_t,\mathbf{r})],
\end{split}
\end{equation}
\end{widetext}
where $\mathcal{N}$ is the normalization factor accounting for the volume of the unit cell and the number of active electrons, indices $n$, $m$, and $f$ iterate over the electronic bands involved into dynamics, integration is performed over the unit cell, $\mathbf{p}_t$ and $\mathbf{p}$ denote the time-dependent and the field-free, respectively, reciprocal space vectors, $\rho_{n,m}(\mathbf{p},t)$ are the matrix elements of the reciprocal space density matrix, $J^{\mu}_{m,n} (\mathbf{p},\mathbf{r})$ denote the normalized $\mathbf{p}$- and $\mathbf{r}$-resolved expectation values of the operator $\hat{J}^{\mu}\tsb{t}$ between bands $m$ and $n$, and $\mathcal{\hat{F}}_{\mathbf{S}}$ is the Fourier transform operator taken with respect to the scattering vector $\mathbf{S}$. The detailed derivation of the approach permitting such simplifications is presented in Ref.~\cite{yuan2025}.

Since graphene is a two-dimensional material and the applied pump field is polarized along $x$ direction, the induced electron dynamics is predominantly confined to the sample plane and occurs primarily along the field polarization. Under these conditions, orienting the probe beam perpendicular to the sample causes the density-current and current-current contributions in Eq.~(\ref{eq:Pks_rel}) to vanish. Indeed, evaluating the corresponding tensor contractions in Eq.~(\ref{eq:Pks_rel}) shows that only the density-density term survives, thereby making the consideration of magnetic effects unnecessary. Therefore, in the following analysis, we assume that the probe beam lies in the $x z$ plane and is oriented at a $45^{\circ}$ angle with respect to the graphene plane. Furthermore, we assume that the sample plane is parallel to the detector, so that only the in-plane components $S_x$ and $S_y$ of the scattering vector $\mathbf{S}$ contribute to the formation of the diffraction pattern. We set the beam energy to 1~MeV to make the corresponding relativistic effects more pronounced in the diffraction image. 

Under the specified conditions, the relativistic electron diffraction signal generally comprises multiple contributions that couple the electron density and the electron current density components along different spatial directions. However, the geometry of the specified experiment makes it possible to isolate specific contributions to the total diffraction signal, such that it predominantly probes either the electron density dynamics or the electron current dynamics. Considering the scattering along $S_x$ and $S_y$ directions in the reciprocal space, the corresponding diffraction intensities, expressed in units of scattering from a free electron, can be written as
\begin{widetext}
\begin{equation}
\label{eq:Ix}
    I(S_x,S_y=0,t)  =  \frac{1}{\mathcal{N}} \sum_{n,m,f} \int_{\text{unit cell}} d \mathbf{p} ~ \rho_{n,m}(\mathbf{p},t)
          \hat{\mathcal{F}}^*_\mathbf{S}[Q_{f,m}(\mathbf{p}_t,\mathbf{r})]
          \hat{\mathcal{F}}_\mathbf{S}[Q_{f,n}(\mathbf{p}_t,\mathbf{r})],
\end{equation}
and
\begin{equation}
\label{eq:Iy}
\begin{split}
    I(S_x=0,S_y,t)  =  \frac{1}{\mathcal{N}} & \sum_{n,m,f} \int_{\text{unit cell}} d \mathbf{p} ~ \rho_{n,m}(\mathbf{p},t)
    \Bigg[
          \hat{\mathcal{F}}^*_\mathbf{S}[Q_{f,m}(\mathbf{p}_t,\mathbf{r})]
          \hat{\mathcal{F}}_\mathbf{S}[Q_{f,n}(\mathbf{p}_t,\mathbf{r})] \\ 
        - & \frac{\sqrt{2}}{2} \beta \alpha \Big(
             \hat{\mathcal{F}}^*_\mathbf{S}[Q_{f,m}(\mathbf{p}_t,\mathbf{r})] \hat{\mathcal{F}}_\mathbf{S}[J^x_{f,n}(\mathbf{p}_t,\mathbf{r})] 
            + \hat{\mathcal{F}}^*_\mathbf{S}[J^x_{f,m}(\mathbf{p}_t,\mathbf{r})] \hat{\mathcal{F}}_\mathbf{S}[Q_{f,n}(\mathbf{p}_t,\mathbf{r})]
        \Big) \\
        + & \frac{\beta^2\alpha^2}{2}
            \hat{\mathcal{F}}^*_\mathbf{S}[J^x_{f,m}(\mathbf{p}_t,\mathbf{r})] \hat{\mathcal{F}}_\mathbf{S}[J^x_{f,n}(\mathbf{p}_t,\mathbf{r})] 
    \Bigg].
\end{split}
\end{equation}
\end{widetext}
Here, $Q_{f,n}(\mathbf{p},\mathbf{r})$ and $J^x_{m,n}(\mathbf{p},\mathbf{r})$ denote the electron density and the $x$-component of the electron current density, respectively, between ($n \neq m$) and within ($n=m$) the corresponding bands, and we have used the relation $|\bk|/E_{\bk}=\alpha \beta$, with $\beta=|\mathbf{v}|/c$ being the dimensionless electron velocity. As seen from Eq.~(\ref{eq:Ix}), the scattering signal along the $x$ direction contains no contributions from density-current or current-current interactions. The scattering signal along the $y$ direction, Eq.~(\ref{eq:Iy}), includes all coupling terms but involves only the current component along $x$. This behavior of the diffraction signals measured at different scattering positions is a direct consequence of the contraction of the corresponding tensors present in Eq.~(\ref{eq:Pks_rel}).

As already discussed in Sec.~\ref{sec:relel} and evident from Eq.~(\ref{eq:Iy}), the density-current and current-current contributions to the interaction between the electron beam and the target are, relative to the density-density term, suppressed by factors proportional to the fine structure constant $\alpha \approx 1/137$ and $\alpha^2$, respectively. Nevertheless, by tuning the energy of the incoming beam and exploiting the fact that, under appropriate conditions, the Fourier transforms appearing in Eq.~(\ref{eq:Iy}) can selectively enhance specific contributions, one can make the relativistic density-current and current-current interactions comparable to or even larger than that arising from the electrostatic coupling, as we demonstrate below.

\begin{figure}[t]
	\includegraphics[width=8.5cm]{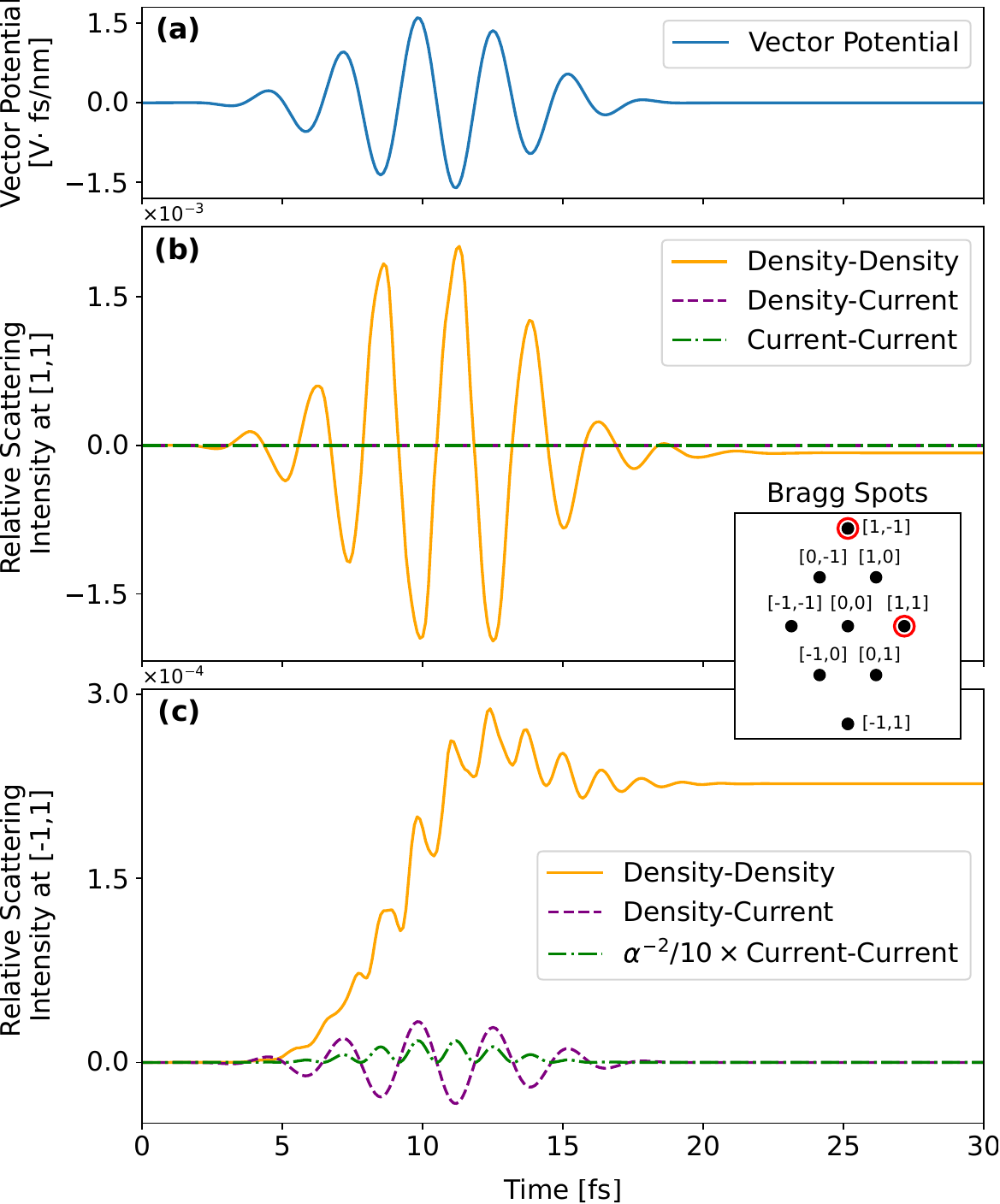}
\caption{Simulated electron diffraction measurements probing the laser-driven dynamics of electron density and electron current density in graphene. (a) Vector potential of the pump field. (b) and (c) Comparison of density-density (orange solid line), density-current (purple dashed line), and current-current (green dash-dotted line) contributions to the time-resolved diffraction signal at the $[1,1]$ and $[1,-1]$ Bragg spots of graphene, respectively. The inset schematically shows the positions of the graphene Bragg spots and highlights those analyzed in panels (b) and (c).}
\label{fig:diffraction}
\end{figure}

We simulated the time-dependent diffraction intensities for the first-order $[1,1]$ (located along $x$ axis) and second-order $[1,-1]$ (located along $y$ axis) Bragg peaks of graphene measuring the dynamics of the electron density and the electron current density illustrated in Fig.~\ref{fig:graphene_dynamics}. Figure~\ref{fig:diffraction} depicts the correspondence between the applied vector potential (shown in panel~(a)) and the observed diffraction signals (shown in panels~(b) and~(c)). As demonstrated in panel~(b) of that figure, the density-density contribution dominates the diffraction signal at the $[1,1]$ Bragg spot, with other effects being completely absent. At the same time, at the $[1,-1]$ Bragg spot, shown in panel (c) of Fig.~\ref{fig:diffraction}, the situation is markedly different, with diffraction oscillations arising from the density-current coupling dominating those due to the density-density interaction. The observed behavior originates from the fact that the location of $[1,-1]$ Bragg spot lies perpendicular to the direction of electron motion in graphene. As a result, electrons scattered into this direction cannot directly exchange momentum with the target electrons, and the density-density contribution just reflects changes in the electron density distribution arising from the population transfer between the valence and conduction bands in graphene. Meanwhile, the electron motion in graphene generates a strong magnetic field that affects the beam electrons, making the dynamics of the electron current density clearly visible in the diffraction signal. The current-current contribution remains small due to the presence of $\alpha^2$ factor, which makes the corresponding signal rather weak. 

Interestingly, the density-density and current-current contributions observed at spot $[1,-1]$ oscillate at twice the frequency of the driving field, while the density-current signal measured at the same spot oscillates at the same frequency. This again reflects the fact that special conditions are required for the electron motion along the $x$-direction in graphene to be observable in a Bragg spot located perpendicular to it. The density-current interaction can be interpreted in terms of its correspondence with the Lorentz force, which shows that a beam electron moving in the field generated by a transverse target current experiences a magnetic deflection perpendicular to both its velocity and the current direction. Consequently, this term encodes the transverse coupling between the direction of the probe beam and the target electron current.

\section{Conclusions}
\label{sec:concl}

In this work, we developed a unified quantum-field-based framework for TR-XRD and TR-UED imaging. We derived a general expression for the scattering probability in which both the free propagation of the probe and its interaction with the target are encoded in matrix elements connecting the initial and scattered states of the beam. As a result, the formalism does not rely on a specific choice of probe or interaction mechanism and can serve as a common starting point for a broad class of scattering problems. 

Using the developed methodology, we recovered the standard structure of TR-XRD and derived the corresponding general theory of TR-UED, thereby placing the two probes on equal footing within a single, consistent theoretical framework. Our analysis clarifies both the similarities and the distinctions between TR-XRD and TR-UED. We found that, under appropriate assumptions, the two techniques become virtually indistinguishable, thereby recovering the equivalence established for their time-independent counterparts.

We established a direct connection between the formulation of TR-UED within the time-dependent perturbation theory and the corresponding description based on the Lippmann--Schwinger equation and the Born series. We demonstrated that the two approaches are fully consistent with one another, and that the resulting expressions for the scattering probabilities are related through simple transformations. In fact, the performed comparison is equivalent to establishing the correspondence between the $S$- and $T$-matrix formulations of scattering theory in the time-dependent regime.

The developed theory of TR-UED extends beyond a simple single-electron picture by retaining a many-body description of the probe beam, thereby enabling a consistent treatment of electron-electron interactions within the beam. We found that the free propagation of the beam enters the corresponding scattering equations via the first order electronic correlation function, which makes it easy to model properties of electron beam and its evolution in space and time.

Considering the relativistic form of the electron-electron interaction Hamiltonian, we incorporated density-current and current-current couplings into TR-UED. We derived a compact, practically useful working equation that includes these contributions alongside the conventional density-density term arising from Coulomb repulsion. This framework extends the physical scope of TR-UED beyond the previously considered regime and enables the exploration of more subtle dynamical features of the target system.

Although the total diffraction signal generally dominated by the density-density interactions, we showed that under suitable conditions current-related contributions can become comparable to, or even exceed, the corresponding density term. We simulated TR-UED signals for laser-driven electron dynamics in graphene and found that, by directing the electron beam at an appropriate angle to the sample, the diffraction becomes sensitive to the electron current dynamics within the target. We thus revealed a regime in which TR-UED can access substantially richer dynamical information than the motion of the electron density alone.

Finally, our approach paves the way for the explicit consideration of specific beam properties such as coherence, duration, polarization, phase stability, potential interference with a reference beam, to name just a few, enabling a more complete and detailed characterization of intricate quantum processes in matter. We hope this work will motivate further studies of both the theoretical foundations and practical applications of TR-XRD and TR-UED in probing ultrafast quantum dynamics in realistic systems.

\begin{acknowledgments}
The computational part of this research is based upon High Performance Computing (HPC) resources supported by the University of Arizona TRIF, UITS, and Research, Innovation, and Impact (RII) and maintained by the UArizona Research Technologies department.
\end{acknowledgments}

\section*{Data availability}
The data that supports the findings of this study are available within the article.

\appendix

\begin{widetext}

\section{Matrix elements of the interaction Hamiltonian for x-ray beam}
\label{app:xray_beam}
In order to describe the interaction of the target with the x-rays consistently, we work in the frame of non-relativistic quantum electrodynamics (QED). We assume the x-ray photon energy be small in comparison to the electron rest energy, so that relativistic QED effects such as pair production may be neglected. Within this regime, the radiation field can be treated as a collection of quantized harmonic oscillators (see, e.g., Ref.~\cite{peskin2019}). We therefore utilize the following Hamiltonian for representing the beam:

\begin{equation}
\label{eq:HEM}
	\hat{H}\tsb{b}\tsp{x-ray} = \sum_{\bk,\lambda} \omega_{\bk} 
	\left(
		\cop \aop + \frac{1}{2}
	\right).
\end{equation}
Here, $\bk$ and $\lambda$ are the wave vector and the polarization index, respectively, of a radiation mode $(\bk,\lambda)$, and $\omega_{\mathbf{k}}=|\mathbf{k}|/\alpha$ is the corresponding photon energy, where $\alpha$ is the fine-structure constant. The operators $\aop$ and $\cop$ are the bosonic annihilation and creation field operators, respectively, which satisfy the following commutation relations:
\begin{eqnarray}
	& \left[ \aop, \cop['] \right]  & = \delta_{\mathbf{k},\mathbf{k}'} \delta_{\lambda,\lambda'}, \\
	& \left[ \aop, \aop['] \right]  & = 0, \\
	& \left[ \cop, \cop['] \right]  & = 0.
\end{eqnarray}

The eigenstates of the Hamiltonian Eq.~(\ref{eq:HEM}) are the bosonic Fock states $|\{n\}\rangle \equiv |n_{\bk_1,\lambda_1}, n_{\bk_2,\lambda_2}, ...\rangle$, which are the photon number states obtained by applying creation operators to the vacuum state $|0\rangle$:
\begin{equation}
	|\psi_j\rangle \equiv |\{n\}\rangle =
		\prod_{\bk,\lambda}
			\frac{(\cop)^{n_{\bk,\lambda}}}{\sqrt{n_{\bk,\lambda}!}}
		|0\rangle.
\end{equation}
Accordingly, $n_{\bk,\lambda}$ specify the number of photons in the mode $(\bk,\lambda)$ and the vacuum state satisfies $\hat a_{\mathbf k,\lambda}|0\rangle=0$. The energies of these states are
\begin{equation}
	E\tsp{b}_j=
		\sum_{\bk,\lambda} \omega_{\bk} \left(n_{\bk,\lambda}+\frac{1}{2}\right),
\end{equation}
so that each photon in mode $(\bk,\lambda)$ contributes an energy $\omega_{\bk}$ on top of the vacuum zero-point energy $1/2 \omega_{\bk}$ per mode.

The action of the creation and annihilation operators on the bosonic Fock states are defined as
\begin{equation}
\label{eq:cop_xray}
    \cop[_j] |\{n\}\rangle = \cop[_j] |...,n_{\bk_j,\lambda_j},...\rangle = 
    \sqrt{n_{\bk_j,\lambda_j}+1} |...,n_{\bk_j,\lambda_j}+1,...\rangle,
\end{equation}
and
\begin{equation}
\label{eq:aop_xray}
    \aop[_i] |\{n\}\rangle = \aop[_i] |...,n_{\bk_i,\lambda_i},...\rangle = 
    \sqrt{n_{\bk_i,\lambda_i}} |...,n_{\bk_i,\lambda_i}-1,...\rangle,
\end{equation}
respectively.

Following from the principle of minimal coupling in the Coulomb gauge and neglecting the contribution from the dispersion correction (see discussion in Refs.~\cite{santra2009,dixit2012} regarding the validity of this approximation), the Hamiltonian describing the interaction between the field and electrons is
\begin{equation}
\label{eq:H_int_Xray}
	\hat{H}^{\text{x-ray}}_{\text{int}} \approx 
		\frac{\alpha^2}{2} 
        \sum_{\sigma} \int d^3 r 
            \hat{\psi}^{\dagger}\tsb{t} (\mathbf{r},\sigma)
			\hat{\mathbf{A}}^2(\mathbf{r})
		\hat{\psi}\tsb{t} (\mathbf{r},\sigma),
\end{equation}
where $\hat{\psi}^{\dagger}\tsb{t} (\mathbf{r},\sigma)$ and $\hat{\psi}\tsb{t} (\mathbf{r},\sigma)$ are the fermionic field operators which create and annihilate, respectively, an electron at
position $\mathbf{r}$ in space with spin projection quantum number $\sigma$. The properties of these operators are discussed in more detail in Appendix~\ref{app:el_beam}. $\hat{\mathbf{A}}(\mathbf{r})$ denotes the vector potential operator, which can be expanded in the plane-wave basis as
\begin{equation}
	\hat{\mathbf{A}}(\mathbf{r}) = \sum_{\bk,\lambda}
	\sqrt{\frac{2\pi}{V\omega_{\bk}\alpha^2}}
		\left\{
			\aop \bm{\varepsilon}_{\mathbf{k},\lambda} e^{i \bk \cdot \mathbf{r}}
			+\cop \bm{\varepsilon}^*_{\mathbf{k},\lambda} e^{-i \bk \cdot \mathbf{r}}
		\right\},
\end{equation}
where $\bm{\varepsilon}_{\bk,\lambda}$ denote the polarization vector. Accordingly, the $\hat{\mathbf{A}}^2(\mathbf{r})$ operator present in Eq.~(\ref{eq:H_int_Xray}) can be written as
\begin{subequations}
\label{eq:A2}
\begin{align}
	\hat{\mathbf{A}}^2(\mathbf{r}) = \frac{2\pi}{V \alpha^2}
	& \sum_{\mathbf{k}_1,\lambda_1} \sum_{\mathbf{k}_2,\lambda_2} 
	\sqrt{\frac{1}{\omega_{\mathbf{k}_1} \omega_{\mathbf{k}_2}}} \\
	\times \Big\{
		    & \aop[_1] \aop[_2]
			(\be[1] \cdot \be[2])
			e^{+i (\mathbf{k}_1 + \mathbf{k}_2) \cdot \mathbf{r}} \\
	+ \quad & \cop[_1] \cop[_2]
			(\be[1]^* \cdot \be[2]^*)
			e^{-i (\mathbf{k}_1 + \mathbf{k}_2) \cdot \mathbf{r}} \\
	+ \quad & \aop[_1] \cop[_2]
			(\be[1] \cdot \be[2]^*)
			e^{+i (\mathbf{k}_1 - \mathbf{k}_2) \cdot \mathbf{r}} \\
	+ \quad & \cop[_1] \aop[_2]
			(\be[1]^* \cdot \be[2])
			e^{-i (\mathbf{k}_1 - \mathbf{k}_2) \cdot \mathbf{r}}					
	\Big\}.
\end{align}
\end{subequations}

Substituting the above expression into Eq.~(\ref{eq:H_int_Xray}), the matrix element $\langle \ns{m} | \hat{H}^{\text{x-ray}}_{\text{int}} | \ns{n} \rangle$ between an initial and the scattered state of the beam can be written as
\begin{equation}
\label{eq:Hint_mel}
\begin{split}
	\langle \ns{m} | \hat{H}^{\text{x-ray}}_{\text{int}} | \ns{n} \rangle = &
		\frac{\pi}{V}	\sum_{\mathbf{k},\lambda}
		\sqrt{\frac{1}{\omega_{\mathbf{k}} \omega_{\ks}}}
			\left(  \bm{\varepsilon}_{\mathbf{k},\lambda} \cdot \bm{\varepsilon}^*_{\ks,\lambda_s} \right)
            \langle \ns{m} | \aop\cop[\tsb{s}] + \cop[\tsb{s}]\aop | \ns{n} \rangle
        \int d^3 r \hat{\rho}(\mathbf{r})
	e^{i (\bk - \bk\tsb{s}) \cdot \mathbf{r}} \\
    = & \frac{2\pi}{V}	\sum_{\mathbf{k},\lambda}
		\sqrt{\frac{1}{\omega_{\mathbf{k}} \omega_{\ks}}}
			\left(  \bm{\varepsilon}_{\mathbf{k},\lambda} \cdot \bm{\varepsilon}^*_{\ks,\lambda_s} \right)
            \langle \ns{m} | \cop[\tsb{s}]\aop | \ns{n} \rangle
        \int d^3 r \hat{\rho}(\mathbf{r})
	e^{i (\bk - \bk\tsb{s}) \cdot \mathbf{r}} \\
    \approx & \frac{2\pi}{V}
		\sqrt{\frac{1}{\omega_{\ki} \omega_{\ks}}}
			\left(  \bm{\varepsilon}_{\ki,\lambda\tsb{in}} \cdot \bm{\varepsilon}^*_{\ks,\lambda_s} \right)
            \sum_{\mathbf{k},\lambda} \langle \ns{m} | \cop[\tsb{s}]\aop | \ns{n} \rangle
        \int d^3 r \hat{\rho}(\mathbf{r})
	e^{i (\bk - \bk\tsb{s}) \cdot \mathbf{r}},
\end{split}
\end{equation}
where $\hat{\rho}(\mathbf{r})=\sum_{\sigma} \hat{\psi}^{\dagger}\tsb{t} (\mathbf{r},\sigma) \hat{\psi}\tsb{t} (\mathbf{r},\sigma)$ denotes the electron density operator of the target. To derive Eq.~(\ref{eq:Hint_mel}), we utilized the fact that the scattered state of interest $|\ns{m}\rangle$ is obtained from the initial one $|\ns{n}\rangle$ by moving one photon from an arbitrary mode $(\bk,\lambda)$ to the scattered mode $(\bk\tsb{s},\lambda\tsb{s})$ that was initially unoccupied:
\begin{equation}
    |\ns{m}\rangle=\frac{\cop[\tsb{s}]\aop}{\sqrt{n_{\bk,\lambda}}} | \ns{n} \rangle \qquad \text{and} \qquad \aop[\tsb{s}] | \ns{n} \rangle = 0,
\end{equation}
where the pre-factor $1/\sqrt{n_{\bk,\lambda}}$ is added to ensure the normalization of $|\ns{m}\rangle$ state. Accordingly, terms~(\ref{eq:A2}b) and~(\ref{eq:A2}c) are all absent in the resulting expression Eq.~(\ref{eq:Hint_mel}) due to the orthogonality of $\aop[_1]\aop[_2] | \ns{n} \rangle$, $\cop[_1]\cop[_2] | \ns{n} \rangle$, and $|\ns{m}\rangle$ states containing different number of particles. Similarly, only those contributions originating from~(\ref{eq:A2}d) and~(\ref{eq:A2}e) survive, that necessarily contain a photon in the mode $(\bk\tsb{s},\lambda\tsb{s})$. Furthermore, we utilized the commutativity, $[\cop[\tsb{s}]\aop]=0$, of the creation and annihilation operators for different photon modes, and relied on the property of the dot product: $\bm{\varepsilon}_{\mathbf{k},\lambda} \cdot \bm{\varepsilon}^*_{\ks,\lambda_s} = \bm{\varepsilon}^*_{\ks,\lambda_s} \cdot \bm{\varepsilon}_{\mathbf{k},\lambda}$. Finally, we assumed that the incident beam is characterized by a mean polarization vector $\bm{\varepsilon}_{\ki,\lambda\tsb{in}}$ and a mean photon energy $\omega_{\ki}$, which allows certain terms to be taken outside of the summation over the field modes.

For the sake of completeness, we also present the required matrix element $\langle \ns{\bar{n}}| \left( \hat{H}\tsp{x-ray}_{\text{int}} \right)^{\dagger} |\ns{m} \rangle$ which can be written as
\begin{equation}
\label{eq:Hint_mel_adj}
\begin{split}
	\langle \ns{\bar{n}}| \left( \hat{H}\tsp{x-ray}_{\text{int}} \right)^{\dagger} |\ns{m} \rangle = 
		\frac{2\pi}{V}	\sum_{\mathbf{k},\lambda}
		\sqrt{\frac{1}{\omega_{\mathbf{k}} \omega_{\ks}}}
			\left(  \bm{\varepsilon}^*_{\mathbf{k},\lambda} \cdot \bm{\varepsilon}_{\ks,\lambda_s} \right)
            \langle \ns{\bar{n}} | \cop\aop[\tsb{s}] | \ns{m} \rangle
        \int d^3 r \hat{\rho}(\mathbf{r})
	e^{-i (\bk - \bk\tsb{s}) \cdot \mathbf{r}} \\
    \approx \frac{2\pi}{V}
		\sqrt{\frac{1}{\omega_{\ki} \omega_{\ks}}}
			\left(  \bm{\varepsilon}^*_{\ki,\lambda\tsb{in}} \cdot \bm{\varepsilon}_{\ks,\lambda_s} \right)
            \sum_{\mathbf{k},\lambda} \langle \ns{\bar{n}} | \cop\aop[\tsb{s}] | \ns{m} \rangle
        \int d^3 r \hat{\rho}(\mathbf{r})
	e^{-i (\bk - \bk\tsb{s}) \cdot \mathbf{r}}.
\end{split}
\end{equation}

In case of transition from $|\psi^{\ki}_i\rangle=|n_{\bk\tsb{in},\lambda\tsb{in}}\rangle$ to $|\psi^{\ks}_f\rangle=|n_{\bk\tsb{in},\lambda\tsb{in}}-1,1_{\bk\tsb{s},\lambda\tsb{s}}\rangle$, i.e. the scattering of one photon to mode $(\bk\tsb{s},\lambda\tsb{s})$ from the continuous beam composed of $n_{\bk\tsb{in},\lambda\tsb{in}}$ photons present in mode $(\bk\tsb{in},\lambda\tsb{in})$, the matrix element becomes
\begin{equation}
\label{eq:Hint_Xray_kin_ks}
	\langle \psi^{\ks}_m | \hat{H}^{\text{x-ray}}_{\text{int}} | \psi^{\ki}_j \rangle = 
		\frac{2\pi}{V}
		\sqrt{\frac{1}{\omega_{\bk\tsb{in}} \omega_{\bk\tsb{s}}}}
		\left(  \be[\tsb{s}]^* \cdot \be[\tsb{in}] \right)
			\sqrt{n_{\bk\tsb{in},\lambda\tsb{in}}}
		\int d^3 r \hat{\rho}(\mathbf{r})
		e^{i (\bk\tsb{in} - \bk\tsb{s}) \cdot \mathbf{r}}.	
\end{equation}
Here, we computed the matrix element
\begin{equation}
    \langle n_{\bk\tsb{in},\lambda\tsb{in}}-1,1_{\bk\tsb{s},\lambda\tsb{s}} | 
        \aop[\tsb{in}]\cop[\tsb{s}] + \cop[\tsb{s}]\aop[\tsb{in}]
    | n_{\bk\tsb{in},\lambda\tsb{in}} \rangle = 2 \sqrt{n_{\bk\tsb{in},\lambda\tsb{in}}}
\end{equation}
explicitly using the rules given in Eqs.~(\ref{eq:aop_xray}) and~(\ref{eq:cop_xray}). The obtained expression, Eq.~(\ref{eq:Hint_Xray_kin_ks}), is the key ingredient required in Eq.~(\ref{eq:Pks_stationary}) to evaluate the probability of elastic scattering of x-rays from a stationary target.

Given the fact that the number states $|\{n\}\rangle$ used in Eq.~(\ref{eq:General_Scattering_Prob}) for the time-dependent scattering probability are the natural eigenstates of the x-ray beam Hamiltonian, Eq.~(\ref{eq:HEM}), the matrix elements $\langle \ns{m}| \hat{U}\tsb{b}(t_f,t_1) \hat{H}_{\text{int}} \hat{U}\tsb{b}(t_1,t_i) |\ns{n} \rangle$ and $\langle \ns{\bar{n}}| \hat{U}^{\dagger}\tsb{b}(t_2,t_i) \hat{H}^{\dagger}_{\text{int}} \hat{U}^{\dagger}\tsb{b}(t_f,t_2) |\ns{m} \rangle$ can be expressed as following:
\begin{equation}
    \langle \ns{m}| \hat{U}\tsb{b}(t_f,t_1) \hat{H}\tsp{x-ray}_{\text{int}} \hat{U}\tsb{b}(t_1,t_i) |\ns{n} \rangle =
        e^{-i E_{\ns{m}}(t_f-t_1)} e^{-i E_{\ns{n}}(t_1-t_i)}
        \langle \ns{m}|
            \hat{H}\tsp{x-ray}_{\text{int}}
        |\ns{n} \rangle,
\end{equation}
and
\begin{equation}
    \langle \ns{\bar{n}}| \hat{U}^{\dagger}\tsb{b}(t_2,t_i) \left( \hat{H}\tsp{x-ray}_{\text{int}} \right)^{\dagger} \hat{U}^{\dagger}\tsb{b}(t_f,t_2) |\ns{m} \rangle =
        e^{i E_{\ns{\bar{n}}}(t_2-t_i)} e^{i E_{\ns{m}}(t_f-t_2)} 
        \langle \ns{\bar{n}}|
            \left( \hat{H}\tsp{x-ray}_{\text{int}} \right)^{\dagger}
        |\ns{m} \rangle.
\end{equation}
These expressions together with the corresponding matrix elements $\langle \ns{m}|\hat{H}\tsp{x-ray}_{\text{int}}|\ns{n} \rangle$ and $\langle \ns{\bar{n}}|\left( \hat{H}\tsp{x-ray}_{\text{int}} \right)^{\dagger}|\ns{m} \rangle$, given by Eqs.~(\ref{eq:Hint_mel}) and~(\ref{eq:Hint_mel_adj}), respectively, are the required ingredients for deriving the probability of x-ray scattering from a time-dependent target.

\section{Matrix elements of the non-relativistic interaction Hamiltonian for electron beam}
\label{app:el_beam}

We consider the non-relativistic electron beam governed by the Hamiltonian (see, e.g., Ref.~\cite{fetter2012})
\begin{equation}
\label{eq:Hb_el}
    \hat{H}\tsb{b}\tsp{el.} = \hat{T} + \hat{V},
\end{equation}
where the kinetic energy term
\begin{equation}
    \hat{T} = \sum_{\sigma} \int d^3 r 
        \hat{\psi}^{\dagger}\tsb{b}(\mathbf{r})
            \left( -\frac{1}{2} \nabla^2 \right)
        \hat{\psi}\tsb{b}(\mathbf{r})
\end{equation}
describes the free propagation of electrons, and the potential energy operator $\hat{V}$ accounts for the electron-electron interactions within the beam. The field operators $\hat{\psi}^{\dagger}\tsb{b}(\mathbf{r},\sigma)$ and $\hat{\psi}\tsb{b}(\mathbf{r},\sigma)$ of the beam are similar to those of the target, $\hat{\psi}^{\dagger}\tsb{t}(\mathbf{r},\sigma)$ and $\hat{\psi}\tsb{t}(\mathbf{r},\sigma)$, introduced in Appendix~\ref{app:xray_beam}. These operators create and annihilate, respectively, an electron at position $\mathbf{r}$ in space with spin projection quantum number $\sigma$. They satisfy the canonical anticommutation relations:
\begin{eqnarray}
	& \left\{ \hat{\psi}\tsb{b}(\mathbf{r},\sigma), \hat{\psi}^{\dagger}\tsb{b}(\mathbf{r}',\sigma') \right\}  & = \delta_{\sigma,\sigma'} \delta^{(3)}({\mathbf{r}-\mathbf{r}'}), \\
	& \left\{ \hat{\psi}\tsb{b}(\mathbf{r},\sigma), \hat{\psi}\tsb{b}(\mathbf{r}',\sigma') \right\}  & = 0, \\
	& \left\{ \hat{\psi}^{\dagger}\tsb{b}(\mathbf{r},\sigma), \hat{\psi}^{\dagger}\tsb{b}(\mathbf{r}',\sigma') \right\}  & = 0.
\end{eqnarray}

We expand the field operators in plane-wave modes within a quantization volume $V$
\begin{subequations}
\label{eq:field_op_exp}
\begin{align}
    \hat{\psi}\tsb{b}(\mathbf{r},\sigma) & =
        \frac{1}{\sqrt{V}} 
            \sum_{\mathbf{k}} e^{i \mathbf{k} \cdot \mathbf{r}} \hat{\mathrm{b}}_{\mathbf{k},\sigma}, \\
    \hat{\psi}^{\dagger}\tsb{b}(\mathbf{r},\sigma) & =
        \frac{1}{\sqrt{V}} 
            \sum_{\mathbf{k}} e^{-\,i \mathbf{k}\cdot \mathbf{r}}\, \hat{\mathrm{b}}_{\mathbf{k},\sigma}^\dagger,
\end{align}
\end{subequations}
thereby expressing the field operators in momentum space, where $\hat{\mathrm{b}}_{\mathbf{k},\sigma}^\dagger$ and $\hat{\mathrm{b}}_{\mathbf{k},\sigma}$ denote the creation and annihilation operators, respectively, of the corresponding mode with momentum $\bk$ and spin $\sigma$. These operators satisfy the following anticommutation relations:
\begin{eqnarray}
    & \left\{ \hat{\mathrm{b}}_{\mathbf{k},\sigma}, \hat{\mathrm{b}}^{\dagger}_{\mathbf{k}',\sigma'} \right\}
        & = \delta_{\mathbf{k},\mathbf{k}'} \delta_{\sigma,\sigma'}, \\
    & \left\{ \hat{\mathrm{b}}_{\mathbf{k},\sigma}, \hat{\mathrm{b}}_{\mathbf{k}',\sigma'} \right\} & = 0, \\
    & \left\{ \hat{\mathrm{b}}^{\dagger}_{\mathbf{k},\sigma}, \hat{\mathrm{b}}^{\dagger}_{\mathbf{k}',\sigma'} \right\} & = 0.
\end{eqnarray}

Substituting the expansions Eq.~(\ref{eq:field_op_exp}) into the free Hamiltonian $\hat{T}$ yields
\begin{equation}
\label{eq:He_beam}
    \hat{T} = 
        \sum_{\mathbf{k},\sigma} \varepsilon_{\mathbf{k}}
        \hat{\mathrm{b}}_{\mathbf{k},\sigma}^{\dagger} \hat{\mathrm{b}}_{\mathbf{k},\sigma},
\end{equation}
where $\varepsilon_{\mathbf{k}} = |\mathbf{k}|^2/2$ is the electron kinetic energy. The eigenstates of the $\hat{T}$ operator are the fermionic Fock states of the form
\begin{equation}
\label{eq:Fock_electron_beam}
    |\{m\}\rangle \equiv
        |m_{\bk_1,\sigma_1}, m_{\bk_2,\sigma_2}, ...\rangle =
        (\hat{\mathrm{b}}^{\dagger}_{\mathbf{k}_1,\sigma_1})^{m_{\mathbf{k}_1,\sigma_1}}
        (\hat{\mathrm{b}}^{\dagger}_{\mathbf{k}_2,\sigma_2})^{m_{\mathbf{k}_2,\sigma_2}}
        \ ...
    |0\rangle,
\end{equation}
where $m_{\mathbf{k}_i,\sigma_i}\in{0,1}$ specifies the occupation of each mode $(\mathbf{k}_i,\sigma_i)$ in accordance with the Pauli exclusion principle, and $|0\rangle$ is the electron vacuum satisfying $\hat{\mathrm{b}}_{\mathbf{k},\sigma}|0\rangle = 0$ for all $\mathbf{k}$ and $\sigma$. In a difference to bosonic case considered in Appendix~\ref{app:xray_beam}, the order of the creation operators in Eq.~(\ref{eq:Fock_electron_beam}) is essential, since they obey fermionic anticommutation relations which enforce the antisymmetric nature of fermionic states.

The action of the creation and annihilation operators on fermionic Fock states is given by
\begin{equation}
    \hat{\mathrm{b}}^{\dagger}_{\bk_i,\sigma_i} | ..., m_{\bk_i,\sigma_i},... \rangle =
    \eta (1-m_{\bk_i,\sigma_i}) | ..., m_{\bk_i,\sigma_i}+1,... \rangle
\end{equation}
and
\begin{equation}
    \hat{\mathrm{b}}_{\bk_i,\sigma_i} | ..., m_{\bk_i,\sigma_i},... \rangle =
    \eta m_{\bk_i,\sigma_i} | ..., m_{\bk_i,\sigma_i}-1,... \rangle.
\end{equation}
These expressions show that a particle can be created only in an unoccupied mode and annihilated only in an occupied one, while the phase factor $\eta=(-1)^{\sum_{j<i} m_{\bk_j,\sigma_j}}$ arises from the fermionic nature of the states.

%
%
%
%

%
%
%

Assuming the kinetic energies of the beam electrons are usually much larger than those of the target, we neglect electron exchange between the target and the beam. We note that electron scattering from the target nuclei is, in general, also present in electron diffraction experiments. In the present work, however, we focus exclusively on the electronic contribution and therefore do not consider the nuclear scattering channel. Therefore, the interaction Hamiltonian between the beam and target electrons takes the form
\begin{equation}
\label{eq:NR_ee_H}
    \hat{H}\tsp{el.}\tsb{int} = 
    \sum_{\sigma,\sigma'}\int d^3 r \int d^3 r'
        \hat{\psi}^{\dagger}\tsb{t}(\mathbf{r},\sigma) \hat{\psi}^{\dagger}\tsb{b}(\mathbf{r}',\sigma')
            \frac{1}{|\mathbf{r}-\mathbf{r}'|}
        \hat{\psi}\tsb{b}(\mathbf{r}',\sigma') \hat{\psi}\tsb{t}(\mathbf{r},\sigma).
\end{equation}
Since the field operators associated with the target and the beam act in separate Hilbert spaces, they commute with one another:
\begin{subequations}
\label{eq:b_t_comm}
\begin{align}
	& \left[ \hat{\psi}\tsb{t}(\mathbf{r},\sigma), \hat{\psi}\tsb{b}(\mathbf{r}',\sigma') \right]  = 0, \\
	& \left[ \hat{\psi}^{\dagger}\tsb{t}(\mathbf{r},\sigma), \hat{\psi}^{\dagger}\tsb{b}(\mathbf{r}',\sigma') \right]  = 0.
\end{align}
\end{subequations}
Furthermore, since these operators refer to different sets of electrons, the beam and target particles are effectively distinguishable in this framework. Consequently, each interacting pair is counted only once, and no additional symmetry factor of $1/2$ is required in the interaction Hamiltonian Eq.~(\ref{eq:NR_ee_H}).

The matrix element $\langle \ns{m} | \hat{H}^{\text{el.}}_{\text{int}} | \ns{n} \rangle$ between an initial and the scattered state of the beam can be written as
\begin{equation}
    \langle \ns{m} | \hat{H}^{\text{el.}}_{\text{int}} | \ns{n} \rangle = 
	\sum_{\sigma'} \int d^3 r \hat{\rho}(\mathbf{r}) 
        \int d^3 r' \langle \ns{m} |
            \hat{\psi}^{\dagger}\tsb{b}(\mathbf{r}',\sigma')
                \frac{1}{|\mathbf{r}-\mathbf{r}'|}
            \hat{\psi}\tsb{b}(\mathbf{r}',\sigma')
        | \ns{n} \rangle,
\end{equation}
where $\hat{\rho}(\mathbf{r})=\sum_{\sigma} \hat{\psi}^{\dagger}\tsb{t}(\mathbf{r},\sigma) \hat{\psi}\tsb{t}(\mathbf{r},\sigma)$ is the electron density operator of the target, and the rearrangement of field operators is possible due to the commutation relations Eqs.~(\ref{eq:b_t_comm}). Using the expansions of the field operators of the beam in the plane-wave basis, Eqs.~(\ref{eq:field_op_exp}), one obtains
\begin{equation}
\begin{split}
    \langle \ns{m} | \hat{H}^{\text{el.}}_{\text{int}} | \ns{n} \rangle & = 
    \frac{1}{V} \sum_{\bk,\bk'} \sum_{\sigma} 
            \langle \ns{m} | \hat{\mathrm{b}}^{\dagger}_{\bk',\sigma} \hat{\mathrm{b}}_{\bk,\sigma} | \ns{n} \rangle
        \int d^3 r \hat{\rho}(\mathbf{r})
        \int d^3 r' \frac{e^{i (\bk - \bk') \cdot \mathbf{r}'}}{|\mathbf{r}-\mathbf{r}'|} \\
    & = \frac{4\pi}{V} \sum_{\bk,\bk'} \sum_{\sigma} 
        \frac{\langle \ns{m} | \hat{\mathrm{b}}^{\dagger}_{\bk',\sigma} \hat{\mathrm{b}}_{\bk,\sigma} | \ns{n} \rangle}
            {|\bk - \bk'|^2} 
        \int d^3 r \hat{\rho}(\mathbf{r}) e^{i (\bk - \bk') \cdot \mathbf{r}'} \\
    & = \frac{4\pi}{V} 
        \sum_{\bk}
        \frac{\langle \ns{m} | \hat{\mathrm{b}}^{\dagger}_{\ks,\sigma\tsb{s}} \hat{\mathrm{b}}_{\mathbf{k},\sigma\tsb{s}} | \ns{n} \rangle}
            {|\bk - \bk\tsb{s}|^2}
            \int d^3 r \hat{\rho}(\mathbf{r}) e^{i (\bk - \bk\tsb{s}) \cdot \mathbf{r}}.
\end{split}
\end{equation}
Here, we evaluated the Fourier transform of the Coulomb potential explicitly, and utilized the fact that only terms leading to the creation of an electron in $\ks$ mode survive since all other terms will be orthogonal to the number state $|\ns{m}\rangle = \hat{\mathrm{b}}^{\dagger}_{\bk\tsb{s},\sigma\tsb{s}} \hat{\mathrm{b}}_{\bk,\sigma} |\ns{n}\rangle$ describing the scattered beam. Note that, since the same spin quantum number $\sigma$ appears in the beam field operators $\hat{\psi}^{\dagger}\tsb{b}(\mathbf{r},\sigma)$ and $\hat{\psi}\tsb{b}(\mathbf{r},\sigma)$, the resulting expression enforces spin conservation during the scattering process.

In the case of scattering of the continuous beam prepared in the $(\ki,\sigma)$ mode, we effectively deal with a single electron only, since the fermionic nature of the number states, together with the Pauli exclusion principle, restricts the occupation of a given mode to at most one electron. Therefore, the initial and the final states of the beam are $|\psi^{\ki}_i\rangle=|1_{\bk\tsb{in},\sigma}\rangle$ and $|\psi^{\ks}_f\rangle=|1_{\bk\tsb{s},\sigma}\rangle$, respectively, and the corresponding matrix element becomes
\begin{equation}
\label{eq:Hint_el_kin_ks}
     \langle \psi^{\ks}_m | \hat{H}^{\text{el.}}_{\text{int}} | \psi^{\ki}_j \rangle = 
        \frac{4\pi}{V} 
            \frac{1}{|\ki - \ks|^2}
            \int d^3 r \hat{\rho}(\mathbf{r}) e^{i (\bk - \bk\tsb{s}) \cdot \mathbf{r}}.
\end{equation}
Similarly to the case of x-ray scattering, the obtained expression, Eq.~(\ref{eq:Hint_el_kin_ks}), is the key ingredient required in Eq.~(\ref{eq:Pks_stationary}) to evaluate the probability of elastic scattering of electrons from a stationary target.

Let us now consider the matrix elements $\langle \ns{m}| \hat{U}\tsb{b}(t_f,t_1) \hat{H}_{\text{int}} \hat{U}\tsb{b}(t_1,t_i) |\ns{n} \rangle$ and $\langle \ns{\bar{n}}| \hat{U}^{\dagger}\tsb{b}(t_2,t_i) \hat{H}^{\dagger}_{\text{int}} \hat{U}^{\dagger}\tsb{b}(t_f,t_2) |\ns{m} \rangle$ required for the description of the time-dependent scattering. In a difference to x-rays, the number states are not the eigenstates of the electron beam Hamiltonian Eq.~(\ref{eq:Hb_el}), and, therefore, more sophisticated treatment must be used to deal with the corresponding matrix elements. At the first step, we employ the same strategy as above, namely rearrange the field operators, expand them in the plane-wave basis, and evaluate the Fourier transform of the Coulomb potential:
\begin{equation}
\label{eq:UHU_el_full}
\begin{split}
        &\langle\{m\}| 
        \hat{U}_{\mathrm{b}}\left(t_f, t_1\right) 
        \hat{H}_{\mathrm{int}} 
        \hat{U}_{\mathrm{b}}\left(t_1, t_i\right)
        |\{n\}\rangle 
        \\
        =& 
        \sum_{\sigma'} \int d^3 r ~
        \hat{\rho}(\mathbf{r})
        \int d^3 r' ~
        \langle\{m\}|
        \hat{U}_{\mathrm{b}}\left(t_f, t_1\right)
        \hat{\psi}^{\dagger}\tsb{b}(\mathbf{r}',\sigma')
        \frac{1}{|\mathbf{r}-\mathbf{r}'|}
        \hat{\psi}\tsb{b}(\mathbf{r}',\sigma')
        \hat{U}_{\mathrm{b}}\left(t_1, t_i\right)
        |\{n\}\rangle \\
    = & \frac{1}{V}  \sum_{\bk,\bk'}\sum_{\sigma}
    \int d^3 r ~
        \hat{\rho}(\mathbf{r}) 
        \langle\{m\}| 
            \hat{U}_{\mathrm{b}}\left(t_f, t_1\right)
            \hat{\mathrm{b}}^{\dagger}_{\bk',\sigma} 
            \hat{\mathrm{b}}_{\bk,\sigma}
            \hat{U}_{\mathrm{b}}\left(t_1, t_i\right)
        |\{n\}\rangle 
        \int d^3 r' \frac{e^{i (\bk - \bk') \cdot \mathbf{r}'}}{|\mathbf{r}-\mathbf{r}'|} \\
    = & \frac{1}{V}  \sum_{\bk,\bk'}\sum_{\sigma} \frac{4\pi}{|\bk - \bk'|^2} 
        \langle\{m\}| 
            \hat{U}_{\mathrm{b}}\left(t_f, t_1\right)
            \hat{\mathrm{b}}^{\dagger}_{\bk',\sigma} 
            \hat{\mathrm{b}}_{\bk,\sigma}
            \hat{U}_{\mathrm{b}}\left(t_1, t_i\right)
        |\{n\}\rangle
    \int d^3 r ~\hat{\rho}(\mathbf{r})        
        e^{i (\bk - \bk') \cdot \mathbf{r}}.
\end{split}
\end{equation}

To proceed forward, let us discuss the physical meaning of the matrix element $\langle\{m\}| \hat{U}_{\mathrm{b}}\left(t_f, t_1\right) \hat{\mathrm{b}}^{\dagger}_{\bk',\sigma} \hat{\mathrm{b}}_{\bk,\sigma} \hat{U}_{\mathrm{b}}\left(t_1, t_i\right) |\{n\}\rangle$ present in Eq.~(\ref{eq:UHU_el_full}). The initial state, represented by the number state $|\ns{n}\rangle$, does not contain electrons in scattering mode $(\ks,\sigma\tsb{s})$. It propagates forward in time from the electron source to the interaction region by the evolution operator $\hat{U}_{\mathrm{b}}\left(t_1, t_i\right)$. At the interaction region, a certain transition defined by the combined operator $\hat{\mathrm{b}}^{\dagger}_{\bk',\sigma} \hat{\mathrm{b}}_{\bk,\sigma}$ takes place. The obtained scattered beam propagates towards detector with $\hat{U}_{\mathrm{b}}\left(t_f, t_1\right)$ operator where the measurement is performed by comparing the time-dependent state of the beam with the number state $|\ns{m}\rangle = \hat{\mathrm{b}}^{\dagger}_{\bk\tsb{s},\sigma\tsb{s}} \hat{\mathrm{b}}_{\bk,\sigma} |\ns{n}\rangle$. While $\hat{U}_{\mathrm{b}}\left(t_1, t_i\right)$ is not an eigenoperator for the utilized number states, indicating that transitions between different momentum modes are expected to occur, these transitions are unlikely to change the direction of the beam propagation substantially. Therefore, the only physically relevant way to obtain the scattered state $|\ns{m}\rangle$ from the initial one $|\ns{n}\rangle$ is to request that the transition driven by the $\hat{\mathrm{b}}^{\dagger}_{\bk',\sigma} \hat{\mathrm{b}}_{\bk,\sigma}$ operator necessarily takes place to $(\ks,\sigma\tsb{s})$ mode, thus leading to $\hat{\mathrm{b}}^{\dagger}_{\bk',\sigma} \hat{\mathrm{b}}_{\bk,\sigma} = \hat{\mathrm{b}}^{\dagger}_{\bk\tsb{s},\sigma\tsb{s}} \hat{\mathrm{b}}_{\bk,\sigma}$ relation. Accordingly, without a loss of generality, Eq.~(\ref{eq:UHU_el_full}) can be simplified as
\begin{equation}
\label{eq:UHU_el_ks}
\begin{split}
    \langle\{m\}| 
    &    \hat{U}_{\mathrm{b}}\left(t_f, t_1\right) 
        \hat{H}_{\mathrm{int}} 
        \hat{U}_{\mathrm{b}}\left(t_1, t_i\right)
    |\{n\}\rangle \\
    = & \frac{1}{V}  \sum_{\bk}\sum_{\sigma} \frac{4\pi}{|\bk - \ks|^2} 
        \langle\{m\}| 
            \hat{U}_{\mathrm{b}}\left(t_f, t_1\right)
            \hat{\mathrm{b}}^{\dagger}_{\ks,\sigma\tsb{s}} 
            \hat{\mathrm{b}}_{\bk,\sigma}
            \hat{U}_{\mathrm{b}}\left(t_1, t_i\right)
        |\{n\}\rangle
    \int d^3 r ~
        \hat{\rho}(\mathbf{r})        
        e^{i (\bk - \ks) \cdot \mathbf{r}},        
\end{split}
\end{equation}
since all other matrix elements in summation over $\bk$ will neccesseraly be zeros.

Let us now consider $\hat{\mathrm{b}}_{\ks,\sigma\tsb{s}} \hat{U}^{\dagger}_{\mathrm{b}}\left(t_f, t_1\right) |\ns{m}\rangle$ term, which is the Hermitian conjugate of $\langle \ns{m}| \hat{U}_{\mathrm{b}}\left(t_f, t_1\right) \hat{\mathrm{b}}^{\dagger}_{\ks,\sigma\tsb{s}}$ present in Eq.~(\ref{eq:UHU_el_ks}). As we already discussed, the number state $|\ns{m}\rangle$ represents a scattered state of the beam measured at the detector with one electron being in $(\ks,\sigma\tsb{s})$ mode. $\hat{U}^{\dagger}_{\mathrm{b}}(t_f, t_1)$ operator propagates this scattered state backward in time from the detector to the interaction region where the transition to $(\ks,\sigma\tsb{s})$ mode happened. Since a single scattered electron is spatially separated from the remaining beam electrons, we can approximate the operator $\hat{\mathrm{b}}_{\ks,\sigma\tsb{s}} \hat{U}^{\dagger}_{\mathrm{b}}\left(t_f, t_1\right)$ as following:
\begin{equation}
\label{eq:noninter_sc_unsc}
    \hat{\mathrm{b}}_{\ks,\sigma\tsb{s}} \hat{U}\tsb{b}^{\dagger}(t_f, t_1) =
    \hat{U}_{\text{b}}^{\dagger}(t_f, t_1) 
        \hat{U}_{\text{b}}(t_f, t_1)    
            \hat{\mathrm{b}}_{\ks,\sigma\tsb{s}}  
        \hat{U}_{\text{b}}^{\dagger}(t_f, t_1)
    \approx
    \hat{U}_{\text{b}}^{\dagger}(t_f, t_1) 
    \hat{\mathrm{b}}_{\ks,\sigma\tsb{s}} 
    e^{i \varepsilon_{\ks}(t_f-t_1)}.
\end{equation}
The latter can be obtained from considering the Heisenberg representation, $\hat{\mathrm{b}}_{\ks,\sigma\tsb{s}}(t) = \hat{U}_{\text{b}}(t) \hat{\mathrm{b}}_{\ks,\sigma\tsb{s}} \hat{U}_{\text{b}}^{\dagger}(t)$, of $\hat{\mathrm{b}}_{\ks,\sigma\tsb{s}}$ operator which satisfies the equation of motion
\begin{equation}
    i\frac{d}{dt}\hat{\mathrm{b}}_{\ks,\sigma\tsb{s}}(t) =
    \big[\hat{\mathrm{b}}_{\ks,\sigma\tsb{s}}(t), \hat{T} + \hat{V}\big].
\end{equation}
Since the commutator $[\hat{\mathrm{b}}_{\ks,\sigma\tsb{s}}(t),\hat{T}]=\varepsilon_{\ks} \hat{\mathrm{b}}_{\ks,\sigma\tsb{s}}(t)$ and we assumed $[\hat{\mathrm{b}}_{\ks,\sigma\tsb{s}}(t),\hat{V}]\approx 0$, i.e. we neglected the interaction of the scattered electron with the rest of the beam, one obtains $\hat{\mathrm{b}}_{\ks,\sigma\tsb{s}}(t) \approx e^{i \varepsilon_{\ks} t} \hat{\mathrm{b}}_{\ks,\sigma\tsb{s}}$. Therefore, we can write
\begin{equation}
\label{eq:Coulomb_Matrix_Element}
\begin{split}
    \langle\{m\}| 
    & \hat{U}_{\mathrm{b}}\left(t_f, t_1\right) 
    \hat{H}_{\mathrm{int}} 
    \hat{U}_{\mathrm{b}}\left(t_1, t_i\right)
    |\{n\}\rangle \\
    \approx & 
    \, e^{-i \varepsilon_{\ks}(t_f-t_1)} \frac{1}{V}  \sum_{\bk}\sum_{\sigma} \frac{4\pi}{|\bk-\ks|^2} 
        \langle\{m\}|
            \hat{\mathrm{b}}^{\dagger}_{\ks,\sigma\tsb{s}} 
            \hat{U}_{\mathrm{b}}\left(t_f, t_1\right)
            \hat{\mathrm{b}}_{\bk,\sigma}
            \hat{U}_{\mathrm{b}}\left(t_1, t_i\right)
        |\{n\}\rangle
    \int d^3 r ~\hat{\rho}(\mathbf{r})  
        e^{i (\bk - \ks) \cdot \mathbf{r}} \\
    \approx& 
    \, e^{-i \varepsilon_{\ks}(t_f-t_1)} \frac{1}{V}  
    \frac{4\pi}{|\bk\tsb{in}-\ks|^2}     
    \sum_{\bk,\sigma}
    \langle\{m\}|
        \hat{\mathrm{b}}^{\dagger}_{\ks,\sigma\tsb{s}} 
        \hat{U}_{\mathrm{b}}\left(t_f, t_1\right)
        \hat{\mathrm{b}}_{\bk,\sigma}
        \hat{U}_{\mathrm{b}}\left(t_1, t_i\right)
    |\{n\}\rangle 
    \int d^3 r ~\hat{\rho}(\mathbf{r})    
        e^{i (\bk - \ks) \cdot \mathbf{r}},
\end{split}
\end{equation}
where we also assumed that the incident beam is characterized by a mean wave vector $\ki$, which allows us to take $4\pi/|\ki-\ks|^2$ factor outside of the summation.

Performing similar set of transformations with the matrix element $\langle \ns{\bar{n}}| \hat{U}_{\mathrm{b}}^\dagger\left(t_2, t_i\right) \left( \hat{H}\tsp{el.}_{\text{int}} \right)^{\dagger} \hat{U}_{\mathrm{b}}^\dagger\left(t_f, t_2\right) |\ns{m} \rangle$, we obtain
\begin{equation}
    \begin{split}
        \langle \ns{\bar{n}}|
        &   \hat{U}_{\mathrm{b}}^\dagger\left(t_2, t_i\right)  
            \left( \hat{H}\tsp{el.}_{\text{int}} \right)^{\dagger} 
            \hat{U}_{\mathrm{b}}^\dagger\left(t_f, t_2\right)
        |\ns{m} \rangle \\
        \approx &        
        \, e^{i \varepsilon_{\ks}(t_f-t_2)} \frac{1}{V}
        \frac{4\pi}{|\ks - \bk\tsb{in}|^2}         
        \sum_{\bk,\sigma}
        \langle\{\bar{n}\}| 
            \hat{U}_{\mathrm{b}}^\dagger\left(t_2, t_i\right) 
            \hat{\mathrm{b}}^{\dagger}_{\bk,\sigma}
            \hat{U}_{\mathrm{b}}^\dagger\left(t_f, t_2\right)
            \hat{\mathrm{b}}_{\ks,\sigma\tsb{s}} 
        |\ns{m}\rangle
        \int d^3 r ~
            \hat{\rho}(\mathbf{r})        
            e^{-i (\bk - \ks)  \cdot \mathbf{r}}.
    \end{split}
\end{equation}

\section{Matrix elements of the relativistic interaction Hamiltonian for electron beam}
\label{app:rel_beam}

In principle, an existing approach permitting a convenient and physically transparent framework for incorporating relativistic corrections into electron-electron interactions while avoiding an explicit treatment of dynamical photons is the well-known Breit Hamiltonian~\cite{breit1932}. Originally, the Breit Hamiltonian is formulated in first quantization form as an effective two-body operator acting on electronic wavefunctions, explicitly dealing with coordinates, momenta, and spin degrees of freedom of the interacting electrons. Here, we use field-theoretic form of the Breit Hamiltonian recasting it in terms of fermionic creation and annihilation operators via connection to the corresponding electron current operators. This formulation allows the necessary interactions to be expressed naturally within Fock space and makes it straightforward to combine with the number-state representation used in the main text. We adopt natural units in this section to align the derivations with standard quantum field theory conventions (see, e.g., Ref.~\cite{peskin2019}). 

Eliminating photons from canonical QED Hamiltonian that describes the coupling between fermionic and photonic degrees of freedom, one can represent the effective Breit interaction Hamiltonian as following
\begin{equation}
\label{eq:H_int_diff}
    \hat H\tsp{B}_{\mathrm{int}} =
    \bar{e}^2 \int d^3r ~ \int d^3r^\prime ~
    \hat{J}\tsb{t}^\mu(\mathbf r)
    D_{\mu\nu}(\mathbf r,\mathbf r^\prime)
    \hat{J}\tsb{b}^\nu(\mathbf r^\prime),
\end{equation}
where $\bar{e}$ is the electron charge and $\hat J^\mu(\mathbf r)=\hat{\bar\Psi}(\mathbf r)\gamma^\mu\hat\Psi(\mathbf r)$ is the Dirac four-current operator, which encodes both the charge density and the probability flow of the particle, with $\gamma^\mu$ denoting the Dirac gamma matrices, and $\hat{\bar\Psi}(\mathbf r)$ and $\hat\Psi(\mathbf r)$ Dirac field operators. Similarly to Eq.~(\ref{eq:NR_ee_H}), the factor $1/2$ is not needed in this expression since we assume no electron exchange between the target and the beam and thus make the corresponding current operators $\hat{J}\tsb{t}^\mu(\mathbf r)$ and $\hat{J}\tsb{b}^\nu(\mathbf r^\prime)$ distinguishable. We restrict the consideration of Dirac fields to the electronic sector and neglect positronic contributions, so that the corresponding field operators are defined as
\begin{subequations}
\label{eq:Dirac_field}
\begin{align}
    \hat{\Psi}(\mathbf{r}) & =\frac{1}{\sqrt{V}}\sum_{\mathbf{k}} \frac{1}{\sqrt{2 E_{\mathbf{k}}}} 
        \sum_{\sigma = \pm 1 / 2}\hat{\mathrm{b}}_{\bk, \sigma} u(k, \sigma) e^{i \mathbf{k} \cdot \mathbf{r}}, \\
    \hat{\bar{\Psi}}(\mathbf{r}) & =\frac{1}{\sqrt{V}}\sum_{\mathbf{k}} \frac{1}{\sqrt{2 E_{\mathbf{k}}}} 
        \sum_{\sigma= \pm 1 / 2}\hat{\mathrm{b}}_{\bk, \sigma}^{\dagger} \bar{u}(k, \sigma) e^{-i \mathbf{k} \cdot \mathbf{r}},
\end{align}
\end{subequations}
where $\bar{u}(k, \sigma)$ and $u(k, \sigma)$ are the Dirac spinors for an electron in $(k,\sigma)$ mode, $k=(k^0,\mathbf{k})$ denote the four-momentum with $k^0=E_{\mathbf{k}}=\sqrt{\bk^2+m^2}$ being the relativistic electron energy, and other quantities have been introduced before in Appendix~\ref{app:el_beam}.

A central quantity in Eq.~(\ref{eq:H_int_diff}) is the effective interaction kernel derived from the Feynman propagator of the photon field:
\begin{equation}
\label{eq:FPP}
    i D_{\mu\nu}(\mathbf r,\mathbf r^\prime) = 
        \int dt \int dt' \ 
        \langle 0|
            \operatorname{T}
                \hat A_{\mu} (t,\mathbf r)
                \hat A_{\nu} (t',\mathbf r')
        |0\rangle,
\end{equation}
where $\hat A_{\mu} (t,\mathbf r)$ is the electromagnetic four-potential operator and $\operatorname{T}$ denotes time ordering. In the Coulomb gauge, the time-integrated components of the interaction kernel can be expressed as (see, e.g., Refs.~\cite{ferenc2022,jentschura2022})
\begin{equation}
\label{eq:Dmunu}
\begin{split}
    D_{00}(\mathbf r,\mathbf r^\prime) & =
        \int \frac{d^3 \mathbf{k}}{(2\pi)^3} \frac{1}{\bk^2} e^{i\bk \cdot (\br-\br')}, \\
    D_{0\nu} &= D_{\mu0}=0,\\
    D_{\mu>0,\nu>0}(\mathbf r,\mathbf r^\prime) &= \int \frac{d^3 \mathbf{k}}{(2\pi)^3} \frac{1}{\bk^2} 
        \left( \frac{k_{\mu} k_{\nu}}{\bk^2} - \delta_{\mu\nu} \right) e^{i\bk \cdot (\br-\br')},
\end{split}
\end{equation}
where the corresponding Fourier transforms can, in principle, be evaluated explicitly, but are left in this form for the sake of simplifying the subsequent analysis. The expressions given by Eqs.~(\ref{eq:Dmunu}) arise from the zero-frequency, also called non-retardation, approximation, which enforces instantaneous interactions between the fermionic fields, effectively projecting the full spacetime propagator onto its zero-frequency component. As a result, retardation effects are neglected, and the interaction reduces to a spatial kernel which depends only on the coordinates $\mathbf r$ and $\mathbf r'$ of the interacting electrons.

We can write the matrix element $\langle\{m\}| \hat{U}_{\mathrm{b}}\left(t_f, t_1\right) \hat{H}_{\mathrm{int}} \hat{U}_{\mathrm{b}}\left(t_1, t_i\right)|\{n\}\rangle$ between an initial and the scattered state of the beam as
\begin{equation}
\begin{split}
    &\langle\{m\}| \hat{U}_{\mathrm{b}}\left(t_f, t_1\right) \hat{H}\tsp{B}_{\mathrm{int}} \hat{U}_{\mathrm{b}}\left(t_1, t_i\right)|\{n\}\rangle
    \\
    = &
    \ \bar{e}^2
    \int d^3 r \int d^3 r^\prime ~
    \hat{J}_{\text{t}}^{\mu}(\mathbf{r})
    D_{\mu \nu}(\mathbf{r},\mathbf{r}^\prime)
    \langle\{m\}| 
    \hat{U}_{\mathrm{b}}\left(t_f, t_1\right)     \hat{J}_{\text{b}}^{\nu}(\mathbf{r}^\prime) \hat{U}_{\mathrm{b}}\left(t_1, t_i\right)
    |\{n\}\rangle \\
    = & \ \frac{\bar{e}^2}{2V}
    \int d^3 r \int d^3 r^\prime ~
    \hat{J}_{\text{t}}^{\mu}(\mathbf{r})
    D_{\mu \nu}(\mathbf{r},\mathbf{r}^\prime)
    \sum_{\bk,\sigma} 
    \sum_{\bk^\prime,\sigma^\prime}
    \frac{e^{i(\bk - \bk^\prime)\cdot \mathbf{r}^\prime}}
        {\sqrt{E_{\bk} E_{\bk^\prime}}}
    \bar{u}(k',\sigma')
    \gamma^\nu 
    u(k,\sigma)
    \langle\{m\}| 
    \hat{U}_{\mathrm{b}}\left(t_f, t_1\right)  
    \hat{\mathrm{b}}_{\bk',\sigma'} ^\dagger 
    \hat{\mathrm{b}}_{\bk,\sigma}  
    \hat{U}_{\mathrm{b}}\left(t_1, t_i\right)
    |\{n\}\rangle,
\end{split}
\end{equation}
where we expanded the $\hat{J}_{\text{b}}^{\nu}(\mathbf{r}^\prime)$ using the expressions Eqs.~(\ref{eq:Dirac_field}). Noting that the incident beam does not contain the mode $(\ks,\sigma\tsb{s})$ prior to the scattering event (see the discussion in the paragraph preceding Eq.~(\ref{eq:UHU_el_ks})) and neglecting interactions between the scattered and unscattered electrons in a similar way we did in Eq.~(\ref{eq:noninter_sc_unsc}), we approximate and rearrange the corresponding matrix element as
\begin{equation}
\label{eq:Matrix_current}
\begin{split}
    \langle\{m\}| 
        \hat{U}_{\mathrm{b}} & \left(t_f, t_1\right) 
        \hat{H}\tsp{B}_{\mathrm{int}} 
        \hat{U}_{\mathrm{b}}\left(t_1, t_i\right)
    |\{n\}\rangle \\
    = \frac{\bar{e}^2}{2V} \int d^3 r & \int d^3 r^\prime ~
    \hat{J}_{\text{t}}^{\mu}(\mathbf{r})
    D_{\mu \nu}(\mathbf{r},\mathbf{r}^\prime)
    \sum_{\bk,\sigma}
    \frac{e^{i(\bk - \bk\tsb{s}) \cdot \mathbf{r}^\prime}}
    {\sqrt{E_{\bk} E_{\bk\tsb{s}}}}
    \bar{u}(k\tsb{s},\sigma\tsb{s})
    \gamma^\nu 
    u(k,\sigma)
    \langle\{m\}|
        \hat{U}_{\mathrm{b}}\left(t_f, t_1\right)
        \hat{\mathrm{b}}_{\bk\tsb{s},\sigma\tsb{s}} ^\dagger
        \hat{\mathrm{b}}_{\bk,\sigma}  
        \hat{U}_{\mathrm{b}}\left(t_1, t_i\right)
    |\{n\}\rangle \\
    & \approx e^{-i E_{\ks}(t_f-t_1)} \frac{\bar{e}^2}{2V}
        \int d^3 r \int d^3 r^\prime ~
        \hat{J}_{\text{t}}^{\mu}(\mathbf{r})
        D_{\mu \nu}(\mathbf{r},\mathbf{r}^\prime) \\
    & \qquad \qquad \times
    \sum_{\mathbf{k},\sigma}
    \frac{e^{i(\mathbf{k} - \ks)\cdot\mathbf{r}^\prime} }{\sqrt{E_{\bk} E_{\ks}}}
    \bar{u}(k\tsb{s},\sigma\tsb{s})
    \gamma^\nu 
    u(k,\sigma)
    \langle\{m\}|
        \hat{\mathrm{b}}_{\ks,\sigma\tsb{s}} ^\dagger 
        \hat{U}_{\mathrm{b}}\left(t_f, t_1\right)  
        \hat{\mathrm{b}}_{\mathbf{k},\sigma}  
        \hat{U}_{\mathrm{b}}\left(t_1, t_i\right)
    |\{n\}\rangle \\
    & = e^{-i E_{\ks}(t_f-t_1)} \frac{\bar{e}^2}{2V}
    \sum_{\mathbf{k},\sigma}
    \frac{1}{\sqrt{E_{\bk} E_{\ks}}}
        \int d^3 r ~
            \hat{J}_{\text{t}}^{\mu}(\mathbf{r})
        \left[
        \int d^3 r^\prime
            D_{\mu \nu}(\mathbf{r},\mathbf{r}^\prime) 
            e^{i(\bk - \ks)\cdot\mathbf{r}^\prime} 
        \right] \\
    & \qquad \qquad \times
    \bar{u}(k\tsb{s},\sigma\tsb{s})
    \gamma^\nu 
    u(k,\sigma)
    \langle\{m\}|
        \hat{\mathrm{b}}_{\ks,\sigma\tsb{s}} ^\dagger 
        \hat{U}_{\mathrm{b}}\left(t_f, t_1\right)  
        \hat{\mathrm{b}}_{\mathbf{k},\sigma}
        \hat{U}_{\mathrm{b}}\left(t_1, t_i\right)
    |\{n\}\rangle.
\end{split}
\end{equation}
Using the definition of $D_{\mu \nu}(\mathbf{r},\mathbf{r}^\prime)$ tensor, Eqs.~(\ref{eq:Dmunu}), the corresponding integral in Eq.~(\ref{eq:Matrix_current}) can be computed explicitly:
\begin{equation}
\label{eq:Dmunu_Q}
    \int d^3 r^\prime
        D_{\mu \nu}(\mathbf{r},\mathbf{r}^\prime)
        e^{ i \mathbf{S}\cdot\mathbf{r}^\prime}
    = \frac{e^{i \mathbf{S} \cdot \br}}{|\mathbf{S}|^2} \widetilde D_{\mu\nu}(\mathbf{S}) 
    = \frac{e^{i \mathbf{S} \cdot \br}}{|\mathbf{S}|^2}
    \begin{cases}
        1 , & \mu=\nu=0,\\
        \frac{S_{\mu} S_{\nu}}{|\mathbf S|^2} - \delta_{\mu\nu},
        & \mu>0,\;\nu>0,\\
        0, & \text{otherwise},
    \end{cases}
\end{equation}    
where $S_{\mu}$ and $S_{\nu}$ are the elements of the four-momentum vector $S=(E_{\mathbf{S}},\mathbf{S})$. Substituting Eq.~(\ref{eq:Dmunu_Q}) to Eq.~(\ref{eq:Matrix_current}), we obtain
\begin{equation}
\label{eq:UHBU}
\begin{split}
    \langle\{m\}| 
        & \hat{U}_{\mathrm{b}}\left(t_f, t_1\right) 
        \hat{H}\tsp{B}_{\mathrm{int}} 
        \hat{U}_{\mathrm{b}}\left(t_1, t_i\right)
    |\{n\}\rangle \\
    & = e^{-i E_{\ks}(t_f-t_1)} \frac{\bar{e}^2}{2V}
    \sum_{\mathbf{k},\sigma}
    \frac{1}{\sqrt{E_{\bk} E_{\ks}}}
    \left[
    \int d^3 r ~
        \hat{J}_{\text{t}}^{\mu}(\mathbf{r}) e^{i(\bk - \ks)\cdot\mathbf{r}}
    \right]
    \frac{\widetilde D_{\mu\nu}(\bk - \ks)}{|\bk-\ks|^2}
    \bar{u}(k\tsb{s},\sigma\tsb{s})
    \gamma^\nu 
    u(k,\sigma) \\
    & \qquad \qquad \times
    \langle\{m\}|
        \hat{\mathrm{b}}_{\ks,\sigma\tsb{s}} ^\dagger 
        \hat{U}_{\mathrm{b}}\left(t_f, t_1\right)  
        \hat{\mathrm{b}}_{\mathbf{k},\sigma}
        \hat{U}_{\mathrm{b}}\left(t_1, t_i\right)
    |\{n\}\rangle \\
    & \approx e^{-i E_{\ks}(t_f-t_1)} \frac{\bar{e}^2}{2V}
    \sum_{\mathbf{k},\sigma}
    \frac{1}{\sqrt{E_{\ki} E_{\ks}}}
    \left[
    \int d^3 r ~
        \hat{J}_{\text{t}}^{\mu}(\mathbf{r}) e^{i(\bk - \ks)\cdot\mathbf{r}}
    \right]
    \frac{\widetilde D_{\mu\nu}(\ki - \ks)}{|\ki-\ks|^2}
    \bar{u}(k\tsb{s},\sigma\tsb{s})
    \gamma^\nu 
    u(k\tsb{in},\sigma) \\
    & \qquad \qquad \times
    \langle\{m\}|
        \hat{\mathrm{b}}_{\ks,\sigma\tsb{s}} ^\dagger 
        \hat{U}_{\mathrm{b}}\left(t_f, t_1\right)  
        \hat{\mathrm{b}}_{\mathbf{k},\sigma}
        \hat{U}_{\mathrm{b}}\left(t_1, t_i\right)
    |\{n\}\rangle \\
    & = e^{-i E_{\ks}(t_f-t_1)} \frac{\bar{e}^2}{2V}
    \frac{\widetilde D_{\mu\nu}(\ki - \ks)}{|\ki-\ks|^2} \frac{1}{\sqrt{E_{\ki} E_{\ks}}}
    \sum_{\sigma}
    \bar{u}(k\tsb{s},\sigma\tsb{s})
    \gamma^\nu 
    u(k\tsb{in},\sigma) \\
    & \qquad \qquad \times \sum_{\mathbf{k}}
    \langle\{m\}|
        \hat{\mathrm{b}}_{\ks,\sigma\tsb{s}} ^\dagger 
        \hat{U}_{\mathrm{b}}\left(t_f, t_1\right)  
        \hat{\mathrm{b}}_{\mathbf{k},\sigma}
        \hat{U}_{\mathrm{b}}\left(t_1, t_i\right)
    |\{n\}\rangle
    \int d^3 r ~
        \hat{J}_{\text{t}}^{\mu}(\mathbf{r}) e^{i(\bk - \ks)\cdot\mathbf{r}},
\end{split}
\end{equation}
where we assumed that the incident electron beam is characterized by a mean momentum vector $\mathbf{k}_{\text{in}}$, such that $E_{\bk} \approx E_{\ki}$, $u(k,\sigma) \approx u(k\tsb{in},\sigma)$, and $\widetilde D_{\mu\nu}(\bk - \ks) \approx \widetilde D_{\mu\nu}(\ki - \ks)$, which also allows rearrangements of tensors since $\widetilde D_{\mu\nu}(\ki - \ks)$ reduces to a number-valued kernel.


Performing similar set of transformations with the matrix element $\langle \ns{\bar{n}}| \hat{U}_{\mathrm{b}}^\dagger\left(t_2, t_i\right) \left( \hat{H}_{\text{int}} \right)^{\dagger} \hat{U}_{\mathrm{b}}^\dagger\left(t_f, t_2\right) |\ns{m} \rangle$, we can write
\begin{equation}
    \begin{split}
        \langle \ns{\bar{n}}|
        &   \hat{U}_{\mathrm{b}}^\dagger\left(t_2, t_i\right)  
            \left( \hat{H}\tsp{B}_{\text{int}} \right)^{\dagger} 
            \hat{U}_{\mathrm{b}}^\dagger\left(t_f, t_2\right)
        |\ns{m} \rangle \\
        \approx &        
        \, e^{i E_{\ks}(t_f-t_2)} \frac{\bar{e}^2}{2V}
    \frac{\widetilde D_{\mu\nu}(\ki - \ks)}{|\ki-\ks|^2} \frac{1}{\sqrt{E_{\ki} E_{\ks}}}
    \sum_{\sigma}
    \bar{u}(k\tsb{in},\sigma)
    \gamma^\nu 
    u(k\tsb{s},\sigma\tsb{s}) \\
    & \qquad \qquad \times \sum_{\mathbf{k}}
        \langle\{\bar{n}\}| 
            \hat{U}_{\mathrm{b}}^\dagger\left(t_2, t_i\right) 
            \hat{\mathrm{b}}^{\dagger}_{\bk,\sigma}
            \hat{U}_{\mathrm{b}}^\dagger\left(t_f, t_2\right)
            \hat{\mathrm{b}}_{\ks,\sigma\tsb{s}} 
        |\ns{m}\rangle
        \int d^3 r ~
            \hat{J}_{\text{t}}^{\mu}(\mathbf{r})
            e^{-i (\bk - \ks)  \cdot \mathbf{r}}.
    \end{split}
\end{equation}

\section{Contraction of spinor products}
\label{app:contraction}

We denote by $Q = k\tsb{in} - k\tsb{s}$ the four-momentum change between the incoming and the scattered electron. We adopt the covariant normalization $\bar{u}(k,\sigma)u(k,\sigma')=2m\,\delta_{\sigma\sigma'}$. By using the identity $\sum_{\sigma} u(k,\sigma) \bar{u}(k,\sigma) = \slashed{k} + m$, where $\slashed{k}\equiv\gamma^\mu k_\mu$ and $m$ is the electron mass, we can rewrite the spinor part as
\begin{equation}
\label{eq:spinors_1}
    \begin{split}
        &\sum_{\sigma_s}
        \bar{u}(k\tsb{in},\sigma\tsb{2})
        \gamma^\nu
        u(k\tsb{s}, \sigma\tsb{s})
        \bar{u}(k\tsb{s}, \sigma\tsb{s})
        \gamma^\alpha
        u(k\tsb{in},\sigma\tsb{1}) \\
    = &
    \ \bar{u}(k\tsb{in},\sigma\tsb{2})
        \gamma^\nu
        (\slashed{k}\tsb{s}+m)
        \gamma^\alpha
        u(k\tsb{in},\sigma\tsb{1}) \\
    = &
    \ \bar{u}(k\tsb{in},\sigma\tsb{2})
        \gamma^\nu
        (\slashed{k}\tsb{in}+m -\slashed{Q})
        \gamma^\alpha
        u(k\tsb{in},\sigma\tsb{1}) \\
    = &
    \ \bar{u}(k\tsb{in},\sigma\tsb{2})
        \gamma^\nu
        (\slashed{k}\tsb{in}+m)
        \gamma^\alpha
        u(k\tsb{in},\sigma\tsb{1}) 
        -
        \bar{u}(k\tsb{in},\sigma\tsb{2})
        \gamma^\nu
        \slashed{Q}
        \gamma^\alpha
        u(k\tsb{in},\sigma\tsb{1}) \\
    \approx &
    \ \bar{u}(k\tsb{in},\sigma\tsb{2})
        \gamma^\nu
        (\slashed{k}\tsb{in}+m)
        \gamma^\alpha
        u(k\tsb{in},\sigma\tsb{1}),
    \end{split}    
\end{equation}
where at the last step of derivation we utilized the fact that $Q \ll k\tsb{in}$, i.e. assumed small momentum transfer, thus neglecting the second term in the above expression.

Using $\{\gamma^\mu,\gamma^\nu\}=2g^{\mu\nu}$, namely $\slashed{k}\tsb{in}\gamma^\alpha = 2k\tsb{in}^\alpha - \gamma^\alpha\slashed{k}\tsb{in}$, together with the on-shell condition $(\slashed{k}-m)u(k,\sigma)=0$, we obtain
\begin{equation}
    (\slashed{k}\tsb{in}+m)\gamma^\alpha
    u(k\tsb{in},\sigma_1)
    =
    (
    2k\tsb{in}^\alpha
    -
    \gamma^\alpha\slashed{k}\tsb{in}
    +
    m\gamma^\alpha
    )
    u(k\tsb{in},\sigma\tsb{1})
    =
    2k\tsb{in}^\alpha
    u(k\tsb{in},\sigma\tsb{1}).
\end{equation}
Therefore, Eq.~(\ref{eq:spinors_1}) can be transformed as
\begin{equation}
    \bar{u}(k\tsb{in},\sigma\tsb{2})
    \gamma^\nu
    (\slashed{k}\tsb{in}+m)
    \gamma^\alpha
    u(k\tsb{in},\sigma\tsb{1})
    =
    2k\tsb{in}^\alpha\,
    \bar{u}(k\tsb{in},\sigma\tsb{2})
    \gamma^\nu
    u(k\tsb{in},\sigma\tsb{1}).
\end{equation}

To evaluate the remaining spinor bilinear, we use
\begin{equation}
    \{\gamma^\nu,\slashed{k}\}=2k^\nu,
    \qquad
    \slashed{k}u(k,\sigma)=m\,u(k,\sigma),
    \quad \text{and} \quad
    \bar{u}(k,\sigma)\slashed{k}
    =
    m\,\bar{u}(k,\sigma),
\end{equation}
which, together with the covariant normalization introduced earlier, gives
\begin{equation}
    \begin{split}
        2m\,
        \bar{u}(k\tsb{in},\sigma\tsb{2})
        \gamma^\nu
        u(k\tsb{in},\sigma\tsb{1})
        &=
        \bar{u}(k\tsb{in},\sigma\tsb{2})
        (
        \slashed{k}\tsb{in}\gamma^\nu
        +
        \gamma^\nu\slashed{k}\tsb{in}
        )
        u(k\tsb{in},\sigma\tsb{1})
        \\
        &=
        \bar{u}(k\tsb{in},\sigma\tsb{2})
        \{\slashed{k}\tsb{in},\gamma^\nu\}
        u(k\tsb{in},\sigma\tsb{1})
        \\
        &=
        2k\tsb{in}^\nu\,
        \bar{u}(k\tsb{in},\sigma\tsb{2})
        u(k\tsb{in},\sigma\tsb{1})
        \\
        &=
        4m\,k\tsb{in}^\nu\,
        \delta_{\sigma\tsb{2}\sigma\tsb{1}}.
    \end{split}
\end{equation}
Therefore,
\begin{equation}
    \bar{u}(k\tsb{in},\sigma\tsb{2})
    \gamma^\nu
    u(k\tsb{in},\sigma\tsb{1})
    =
    2k\tsb{in}^\nu\,
    \delta_{\sigma\tsb{2}\sigma\tsb{1}},
\end{equation}
and hence
\begin{equation}
    \sum_{\sigma\tsb{s}}
    \bar{u}(k\tsb{in},\sigma\tsb{2})
    \gamma^\nu
    u(k\tsb{s},\sigma\tsb{s})
    \bar{u}(k\tsb{s},\sigma\tsb{s})
    \gamma^\alpha
    u(k\tsb{in},\sigma\tsb{1}) 
    \approx
    4k\tsb{in}^\nu
    k\tsb{in}^\alpha
    \delta_{\sigma\tsb{2}\sigma\tsb{1}}.
\end{equation}

We would like to mention also that if the observable is spin resolved, one obtains a similar result
\begin{equation}
    \bar{u}(k\tsb{in},\sigma\tsb{2})
    \gamma^\nu
    u(k\tsb{s},\sigma\tsb{s})
    \bar{u}(k\tsb{s},\sigma\tsb{s})
    \gamma^\alpha
    u(k\tsb{in},\sigma\tsb{1}) 
    \approx
    4k\tsb{in}^\nu
    k\tsb{in}^\alpha
    \delta_{\sigma\tsb{2}\sigma\tsb{s}} 
    \delta_{\sigma\tsb{1}\sigma\tsb{s}},
\end{equation}
which indicates that the spin of electron beam remains conserved in the scattering experiments.

\end{widetext}

\bibliographystyle{apsrev4-2}
\bibliography{eXray.bib}

\end{document}